\def\papertitle{Modulation Extraction for LFO-driven Audio Effects}
\def\paperauthorA{Christopher Mitcheltree}
\def\paperauthorB{Christian J. Steinmetz}
\def\paperauthorC{Marco Comunità}
\def\paperauthorD{Joshua D. Reiss}

\documentclass[twoside,a4paper]{article}
\usepackage{etoolbox}
\usepackage{dafx_23}
\usepackage{amsmath,amssymb,amsfonts,amsthm}
\usepackage{euscript}
\usepackage[T1]{fontenc}
\usepackage[utf8]{inputenc}
\usepackage{ifpdf}
\usepackage[english]{babel}
\usepackage{caption}
\usepackage{subfig} 
\usepackage{color}
\usepackage{booktabs}
\usepackage{cite}
\usepackage{enumitem}
\input glyphtounicode
\ninept

\newcounter{numauth}
\newcounter{listcnt}
\newcommand\authcnt[1]{\ifdefined#1 \stepcounter{numauth} \fi}

\newcommand\addauth[1]{
\ifdefined#1 
\stepcounter{listcnt}
\ifnum \value{listcnt}<\value{numauth}
\appto\authorslist{, #1}
\else
\appto\authorslist{~and~#1}
\fi
\fi}
\authcnt{\paperauthorB}
\authcnt{\paperauthorC}
\authcnt{\paperauthorD}
\authcnt{\paperauthorE}
\authcnt{\paperauthorF}
\authcnt{\paperauthorG}
\authcnt{\paperauthorH}
\authcnt{\paperauthorI}
\authcnt{\paperauthorJ}
\def\authorslist{\paperauthorA}
\addauth{\paperauthorB}
\addauth{\paperauthorC}
\addauth{\paperauthorD}
\addauth{\paperauthorE}
\addauth{\paperauthorF}
\addauth{\paperauthorG}
\addauth{\paperauthorH}
\addauth{\paperauthorI}
\addauth{\paperauthorJ}

\usepackage{times}

\newif\ifpdf
\ifx\pdfoutput\relax
\else
   \ifcase\pdfoutput
      \pdffalse
   \else
      \pdftrue
\fi

\ifpdf 
  \usepackage[pdftex,
    pdftitle={\papertitle},
    pdfauthor={\authorslist},
    pdfsubject={Proceedings of the 26th International Conference on Digital Audio Effects (DAFx23)},
    colorlinks=false, 
    bookmarksnumbered, 
    pdfstartview=XYZ 
  ]{hyperref}
  \usepackage[pdftex]{graphicx}
\else 
  \usepackage[dvips]{epsfig,graphicx}
  \usepackage[dvips,
    pdftitle={\papertitle},
    pdfauthor={\authorslist},
    pdfsubject={Proceedings of the 26th International Conference on Digital Audio Effects (DAFx23)},
    colorlinks=false, 
    bookmarksnumbered, 
    pdfstartview=XYZ 
  ]{hyperref}
\fi
\usepackage[hypcap=true]{caption}
\title{\papertitle}

\usepackage[dvipsnames]{xcolor}
\newcommand\myshade{70}
\colorlet{mywholecolor}{MidnightBlue}
\hypersetup{
  linkcolor  = mywholecolor!\myshade!black,
  citecolor  = mywholecolor!\myshade!black,
  urlcolor   = mywholecolor!\myshade!black,
  colorlinks = true,
}

\newcommand\linesubsec[1]{\vspace{0.8mm}\noindent\textbf{#1 --- }}
\affiliation{
\paperauthorA \quad\quad
\paperauthorB \quad\quad
\paperauthorC \quad\quad
\paperauthorD}
{Centre for Digital Music, Queen Mary University of London, UK\\
{\tt \{c.mitcheltree, 
     c.j.steinmetz,
    m.comunita,
    joshua.reiss\}@qmul.ac.uk }
}

\begin{document}
\ifpdf 
  \DeclareGraphicsExtensions{.png,.jpg,.pdf}
\else  
  \DeclareGraphicsExtensions{.eps}
\fi


\maketitle

\begin{abstract}
Low frequency oscillator (LFO) driven audio effects such as phaser, flanger, and chorus, modify an input signal using time-varying filters and delays, resulting in characteristic sweeping or widening effects.
It has been shown that these effects can be modeled using neural networks when conditioned with the ground truth LFO signal. 
However, in most cases, the LFO signal is not accessible and measurement from the audio signal is nontrivial, hindering the modeling process.
To address this, we propose a framework capable of extracting arbitrary LFO signals from processed audio across multiple digital audio effects, parameter settings, and instrument configurations.
Since our system imposes no restrictions on the LFO signal shape, we demonstrate its ability to extract quasiperiodic, combined, and distorted modulation signals that are relevant to effect modeling.
Furthermore, we show how coupling the extraction model with a simple processing network enables training of end-to-end black-box models of unseen analog or digital LFO-driven audio effects using only dry and wet audio pairs, overcoming the need to access the audio effect or internal LFO signal.  
We make our code available and provide the trained audio effect models in a real-time VST plugin\footnote{\href{https://christhetree.github.io/mod\_extraction/}{https://christhetree.github.io/mod\_extraction/}}.
\end{abstract}

\section{Introduction}
\label{sec:intro}
\vspace{-4pt}

\begin{figure}
    \centering
    \includegraphics[width=\linewidth,trim={0.1cm 0.1cm 0.0cm 0.2cm},clip]{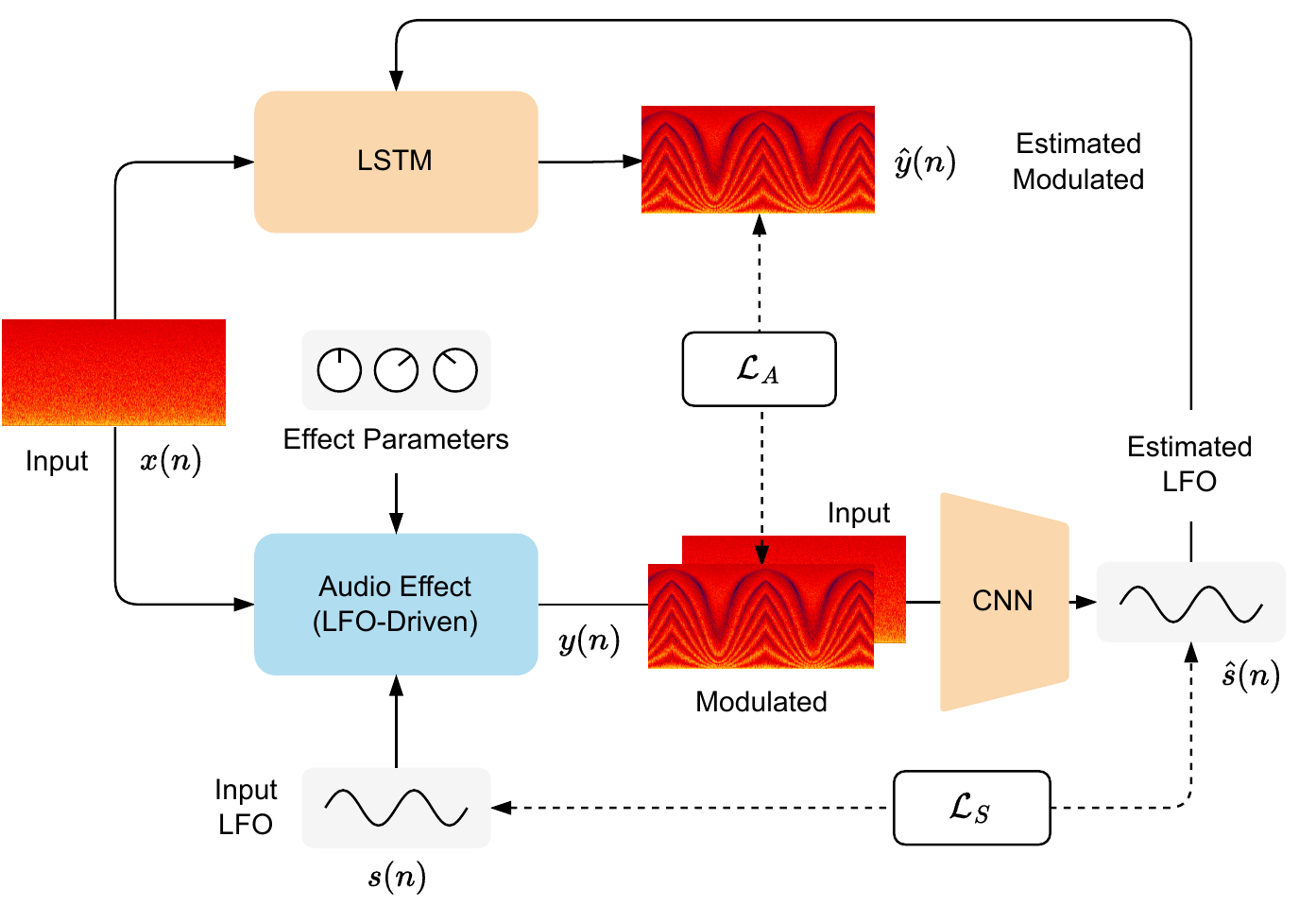}
    \caption{By using the pretrained LFO extraction model (CNN) to analyze input and modulated audio, our proposed system enables training of a black-box neural network model (LSTM) on modulation audio effects without access to the ground truth LFO signal.}
    \label{fig:system_diagram}
\end{figure}

%

In music composition, production, and engineering, audio effects play a key role in altering the sound toward the desired result. 
Modulation effects such as phaser, flanger, and chorus, are part of a broad family of audio processors based on using a modulation signal to modify the spectrum, loudness, or spatial characteristic of the input audio. 
The typical modulation signals adopted are periodic (e.g., sinusoidal, sawtooth, triangular) with a frequency below the audible range (20\,Hz) and are therefore called low frequency oscillators (LFO).
Since oscillators are used to continuously vary the internal parameters of these effects, the exact shape, frequency, and phase of the LFO signal plays a crucial role, affecting the overall timbre and temporal behavior. 
This is especially evident in analog circuits where imperfections and nonlinearities may cause distortion and quasi-periodicity of the oscillation.

Digital emulation of audio effects is an area of active research \cite{pakarinen2009review, yeh2009automated, valimaki2010introduction}, and many methods have been developed to analyze and emulate effect units. Depending on the degree of prior knowledge and reliance on measurement data, these can be divided into white- \cite{huovilainen2005enhanced, parker2011simple, eichas2014physical, mavcak2016simulation}, gray-~\cite{smith1984allpass, kiiski2016time, darabundit2019digital, wright2020neural, wright2021neural, colonel2022reverse}, or black-box~\cite{ramirez2019general, ramirez2020deep, steinmetz2022efficient, comunità2022modelling} approaches.
Most prior work on modulation effects modeling uses complex and time-\\consuming white-box approaches,
obtaining models that are not easily transferable to other designs or LFO-driven effects.
There are also examples of gray-box approaches, which strike a balance between general validity of the block-based model~\cite{smith1984allpass} and emulation quality of a specific unit~\cite{kiiski2016time, darabundit2019digital}. 
However, the modeling capabilities and robustness of such models is limited by the hand-engineered measurement techniques used to extract the LFO signal~\cite{kiiski2016time, darabundit2019digital, wright2020neural, wright2021neural} and assumptions made about the LFO's shape.


Work in \cite{ramirez2019general, ramirez2020deep} proposes recurrent and convolutional neural networks for black-box modeling of time-varying audio effects. Relying only on datasets of dry-wet audio, these are the first end-to-end approaches applied to LFO-driven effects. The method achieves good results with non-causal and non-controllable implementations, but does not explicitly learn the LFO signal and is not evaluated on unseen effects, audio, or LFO shapes.




To address the challenges of modeling a wide range of modulation effects and configurations, we introduce a neural architecture that is trained to extract arbitrary LFO signals from phaser, flanger, and chorus audio effects across varying parameter settings.
By training this model on a dataset of guitar signals with basic phaser, flanger, and chorus implementations, we demonstrate our model achieves:

\begin{itemize}[leftmargin=*]
\setlength\itemsep{0.2em}
    \item Accurate modulation extraction from unseen audio sources.
    \item Extrapolation to complex modulation signals such as quasiperiodic, combined, and distorted LFO signals.
    \item Generalization across effect implementations for unseen analog and digital phaser and flanger effects.
    \item End-to-end causal modeling of analog and digital LFO-driven effects without access to the internal LFO signal.
\end{itemize}










\section{Background}
\subsection{Low Frequency Oscillators}

In 1964 Robert Moog introduced the first transistor-based voltage controlled oscillator (VCO) and voltage controlled amplifier (VCA) designs~\cite{moog1965voltage}. 
These circuits are at the origin of modular synthesizers and later on led to modulation audio effects like phaser, flanger, and chorus. 
While VCOs were used to generate pitched sounds, VCAs were responsible for the envelope of synthesized notes. 
In his designs Moog also included VCOs oscillating at frequencies below 20\,Hz, i.e. LFOs, to modulate other signal parameters (e.g., frequency, phase, amplitude) or processing blocks (e.g., panning, cutoff frequency).
The most common types of modulations stem from periodic waveforms like sine, sawtooth, triangle, or square, but often extend to more complex shapes.

In analog effects~\cite{darabundit2019digital, huovilainen2005enhanced}, non-ideal components can cause distortions from the reference shape as well as deviations that cause quasiperiodicity.
There are also cases, like chorus effects, where random LFO signals are adopted.
Furthermore, with the prevalence of digital emulations and software synthesizers, modulation signals can achieve an even wider diversity than their analog counterparts. 
As a result, the extraction of modulation signals from processed audio has applications beyond virtual analog modeling.

\subsection{Modulation Effects}
Phaser and flanger are examples of modulations affecting the spectrum of a signal, while chorus affects the pitch and timing.

\linesubsec{Phaser} 
Phasing is achieved by using a series of notch or allpass filters~\cite{hartmann1978flanging}. The typical analog implementation uses an even number of first order allpass filters, which have flat magnitude response but phase that varies between $0^{\circ}$ and $-180^{\circ}$. When two filters are connected in series the phase varies back to $0^{\circ}$ and, by mixing the filtered output with the input signal phase cancellations occur at frequencies around the $180^{\circ}$ point.
Altering the center frequency of the filters creates a characteristic sweeping sound.

\linesubsec{Flanger} In a flanger, a delayed copy of the input signal is summed to the dry input itself causing constructive and destructive interference. The delay is periodically modulated but usually kept below $\approx15$\,ms. As a result, it is often perceived as a time-varying comb filter. In contrast to phaser effects, where the frequency distance between notches is kept constant on a logarithmic scale, in flanger effects, the distance changes with the delay value.

 \linesubsec{Chorus} Chorus effects are identical to flangers in implementation, but use multiple delayed and modulated copies of the input signal. Also, by adopting larger delays - around $\approx30$\,ms - the output is perceived as a sum of slightly pitch shifted copies of the input, as when multiple instruments or voices are playing in unison. Therefore, there is not a clearly observable modulation of the spectrum compared to phasers and flangers.



\subsection{Virtual Analog Modeling of Modulation Effects}
Research in virtual analog modeling aims to develop methods that emulate the characteristics and behaviors of a reference unit. These methods can be divided into white-, gray-, or black-box modeling depending on the degree of prior knowledge and type of measurements they rely on.
To create accurate simulations, white-box modeling~\cite{huovilainen2005enhanced, parker2011simple, eichas2014physical, mavcak2016simulation} requires a thorough understanding of the system, and typically employs differential equations to describe its behavior and numerical methods to solve them. Therefore, such methods are often associated with a time consuming design process and computationally demanding and non-transferable implementations.
Circuit analysis together with voltage and current measurements are used to create a state-space model of a phaser effect pedal in \cite{eichas2014physical}, while in \cite{mavcak2016simulation} a similar analysis is used to emulate a bucket brigade delay circuit that is then employed in flanger effect emulation. Phaser, flanger, and chorus are also modeled in \cite{huovilainen2005enhanced}, where the authors discretize the differential equations 
of JFET transistors and transconductance amplifiers used in such effects.


To reduce prior knowledge necessary to model a device, gray-box approaches combine a partially theoretical structure 
with input-output measurements \cite{kiiski2016time, darabundit2019digital, wright2021neural}. However, they still require ad hoc measurement and optimization procedures \cite{kiiski2016time, wright2021neural} and knowledge of the underlying implementation.
A gray-box model of phaser effect pedal is presented in \cite{darabundit2019digital}, where 
nonlinear allpass filter blocks are combined with analysis and measurement of the interaction between light dependent resistors and incandescent lamp optocoupler controlling the LFO.
This work shows how critical the LFO signal can be in shaping the overall sound of a design.
In \cite{kiiski2016time} we have an example of a measurement signal and extraction algorithm specifically designed to capture a phaser's LFO signal. 

A similar measurement is adopted in~\cite{wright2020neural}, 
and the extracted LFO signal is used to condition neural networks trained on phaser and flanger effects. A custom extraction algorithm is implemented: the LFO shape (rectified sine) is observed in the output and given to a least-squares solver. Furthermore, custom training data is required, where the test signal is interposed between samples so that the initial LFO phase can be extracted.
This work is developed further in~\cite{wright2021neural} by improving the measurement technique.


In black-box approaches, minimal knowledge of the system is required and modeling mostly relies on input-output measurements. A major advantage 
is that they simplify the process to collecting adequate data. However, these models often lack interpretability and might entail time-consuming optimizations.
In \cite{ramirez2019general, ramirez2020deep}, we have the only examples of black-box models of time-varying audio effects. 
Neural networks are successfully trained on many modulation effect types. However, these models are non-causal, non-controllable, and have not been tested on unseen LFO shapes or audio signals different from the training data.

\subsection{Effects Recognition and Parameter Estimation}
Beyond effect modeling, there has also been research on recognition of audio effects and effect chains, as well as control values from processed audio. 
Our task of LFO extraction can be viewed as a specific form of audio effect parameter estimation, however, existing works have yet to consider reconstructing the LFO signal itself.
Early works focus on audio effects classification~\cite{stein2010Aautomatic}, while others extend this task to target the identification of specific effect units and their control values~\cite{comunita2021guitar}, including within mixtures~\cite{hinrichs2022convolutional}. 
Recently, work has generalized this task to the complete reconstruction of a graph of audio effects and their parameter values \cite{lee2023blind}.
In \cite{sheng2019feature}, the authors focus on dynamic range compression, and train neural networks to extract ratio, attack, and release times, and total harmonic distortion from a reference signal. 
Extracting information from audio recordings for applications in music production and sound synthesis is still at an early stage, and the work presented here also aims to contribute in these directions. 

\section{Methodology}
We approach the problem of modeling an LFO-driven audio effect in two steps. First, we develop an LFO extraction model which can be trained to reconstruct the modulation signal from dry and wet audio pairs. Then we feed the extracted LFO signal along with the dry audio to an effect model that can be trained to reconstruct the wet audio. Figure \ref{fig:system_diagram} visualizes our approach with a block diagram.

\subsection{LFO Extraction}
\begin{figure}[t!]
    \centering
    \includegraphics[width=0.85\linewidth,trim={0.1cm 0.3cm 0.1cm 0.2cm},clip]{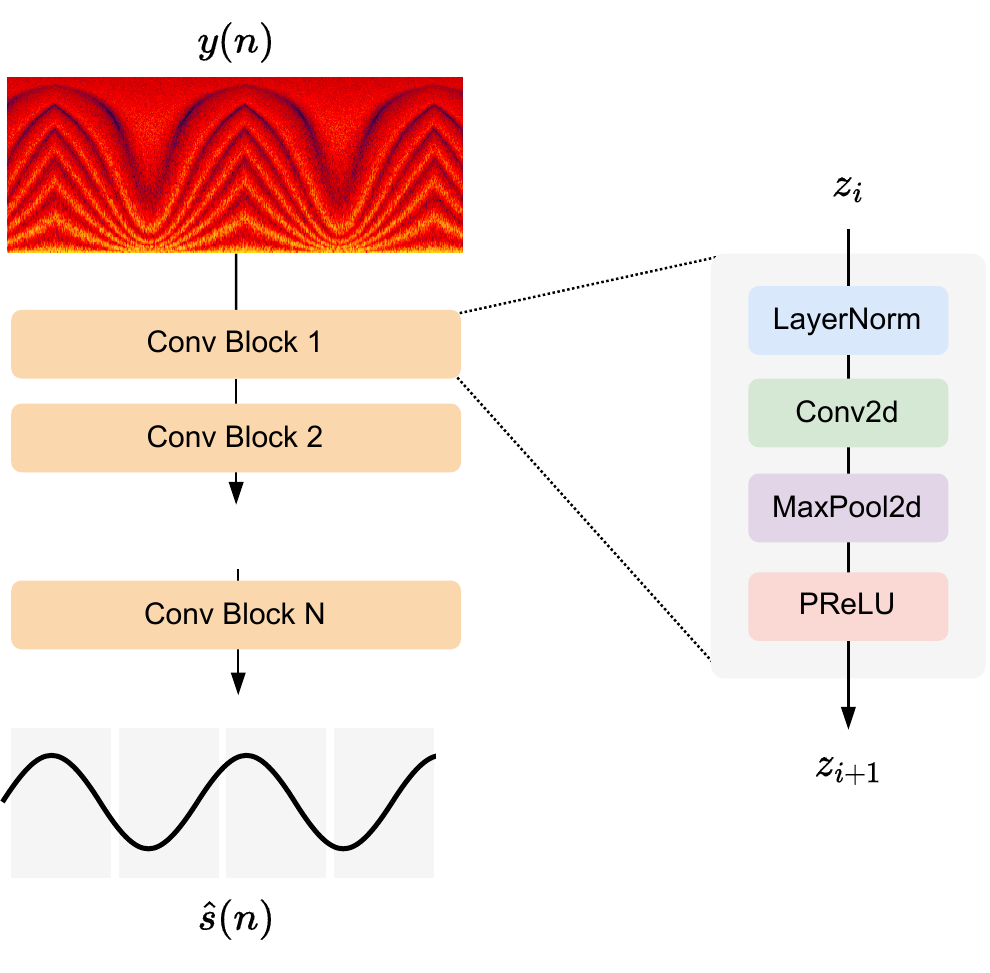}
    \caption{LFO extraction model (LFO-net) diagram.}
    \label{fig:lfo_model}
\end{figure}

The LFO extraction model (LFO-net), shown in Figure \ref{fig:lfo_model}, is a convolutional neural network (CNN) consisting of sequential convolutional blocks. As input it takes a 2-channel Mel spectrogram of the dry and wet audio. Each block consists of LayerNorm~\cite{ba2016layer} across the frequency and time dimensions, a 2D convolution, Max Pooling, and a PReLU activation~\cite{he2015delving}. Feature maps are max-pooled only along the frequency dimension and dilated only along the time dimension, similar to how a 
temporal convolutional network (TCN) operates \cite{lea2017temporal}. As a result, the temporal receptive field of the network grows exponentially with each convolutional block while the frequency resolution decreases exponentially. The final layer of the network is a time-distributed linear layer that estimates the LFO value for the current frame between 0 and 1.

\linesubsec{Training}
LFO-net is trained using the AdamW optimizer
to minimize the $L_1$ error between the reconstructed modulation signal $\hat{s}$ and the ground truth modulation signal $s$, each with $N$ timesteps
\vspace{-0.2cm}
\begin{equation}
\label{eg:l1_loss}
L_1(s, \hat{s}) = \frac{1}{N} \sum_{n=1}^{N} \left| s(n) - \hat{s}(n) \right|
\end{equation}
where $n$ is the time index.
We also include terms for the first-order central difference error
\vspace{-0.2cm}
\begin{equation}
s'(n) = \frac{s(n+1) - s(n-1)}{2}.
\label{eg:fd_loss}
\end{equation}
as well as the $L_1$ error of the second-order central difference, which is defined recursively
\vspace{-0.2cm}
\begin{equation}
s''(n) = \frac{s'({n+1}) - s'({n-1})}{2}.
\label{eg:sd_loss}
\end{equation}
These terms are scaled by $\alpha$, $\beta$, and $\gamma$ respectively and encourage the network to learn smoother modulation signals. 
The complete loss function $\mathcal{L}_S$ can be expressed as follows:
\begin{equation}
\mathcal{L}_S = \alpha L_1 \left(s, \hat{s} \right) + \beta L_1 \left(s',\hat{s}' \right) + \gamma L_1 \left(s'', \hat{s}'' \right)
\label{eg:lfo_loss}
\end{equation}
Based on initial testing, we selected $\alpha=1$, $\beta=5$, and $\gamma=10$ to weigh the different terms.
In addition, SpecAugment~\cite{park2019specaugment} was used for masking both frequency and time dimensions during training to increase robustness.

\begin{figure}[t!]
\centering
\begin{minipage}{.32\linewidth}
    \centering
    \includegraphics[width=1.0\linewidth]{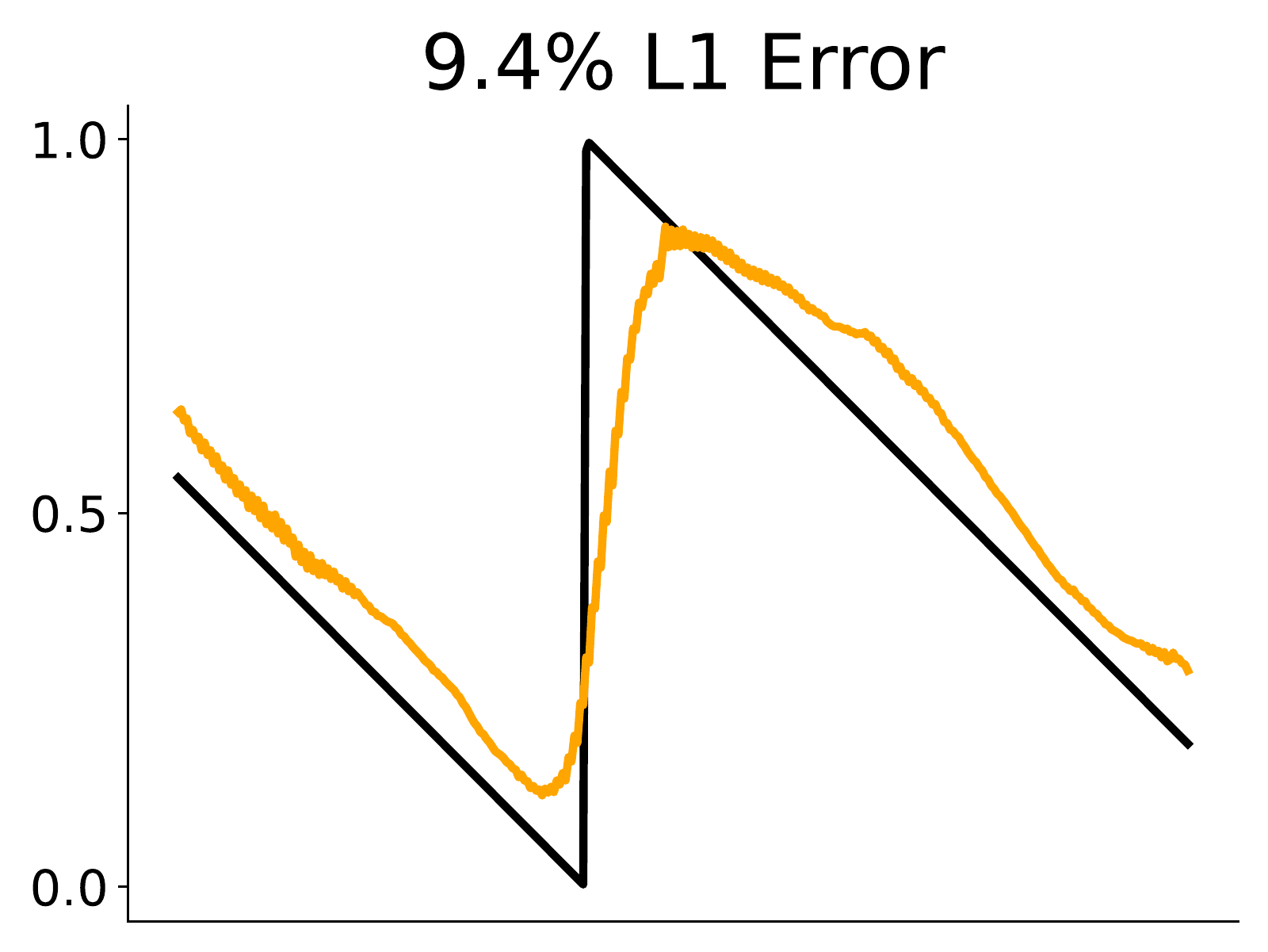}
\end{minipage}
\begin{minipage}{.32\linewidth}
    \centering
    \includegraphics[width=1.0\linewidth]{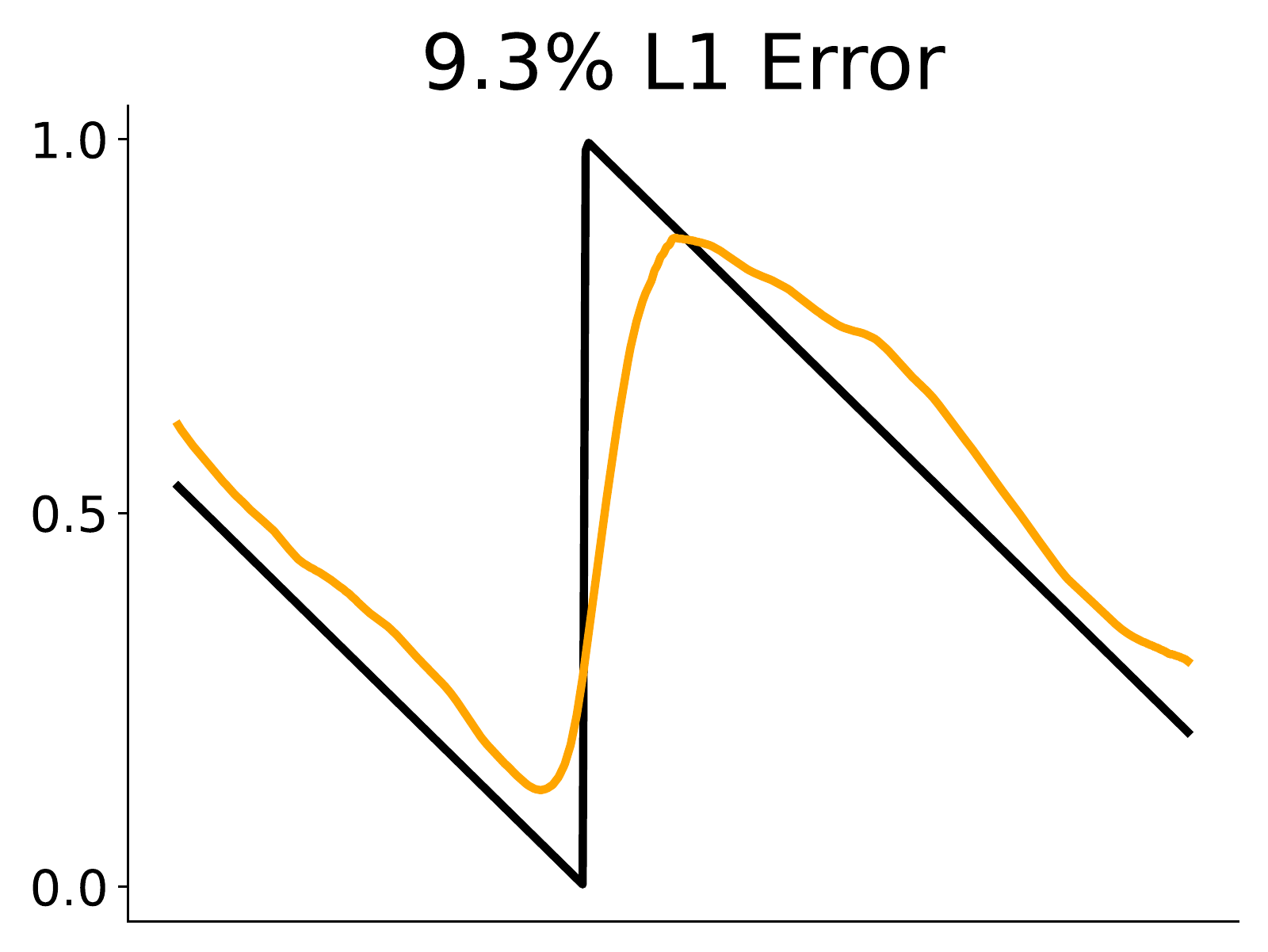}
\end{minipage}
\begin{minipage}{.32\linewidth}
    \centering
    \includegraphics[width=1.0\linewidth]{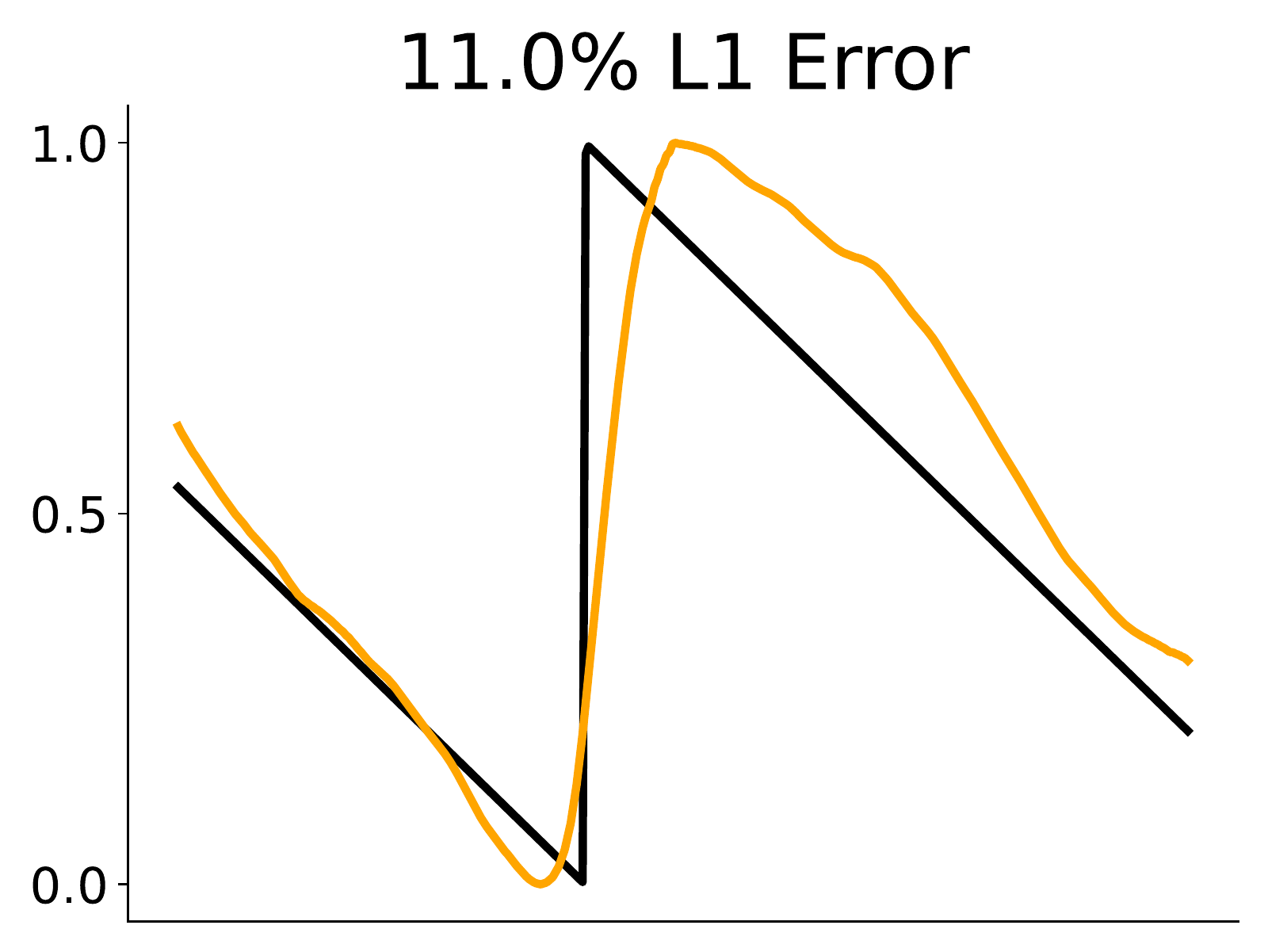}
\end{minipage}
\vspace{-0.1cm}
\caption{Examples of original (left), smoothed (center), and stretched (right) post-processed modulation signals.}
\label{fig:postprocessing}
\vspace{-0.2cm}
\end{figure}

\linesubsec{Post-processing}
\label{sec:postprocessing}
Since LFO-net imposes no restrictions on the shape of the LFO besides being bounded between 0 and 1, the output can appear noisy or irregular. To improve the quality of the extracted LFO signal we introduce three post-processing steps, shown in Figure~\ref{fig:postprocessing}. First, the signal is smoothed with a 4th order moving average filter. This is followed by ``stretching'' of the peaks and troughs so they are equal to 0 and 1, respectively. This is achieved by finding the locations of local minima and maxima, and then linearly interpolating consecutive sections to span from 0 to 1. 
Finally, when training effect models, invalid reconstructed LFO signals where there are too many peaks or troughs or where consecutive peaks or troughs are too close together are thrown out.
This helps stabilize training by only using examples where the estimated LFO is likely to be an accurate prediction of the ground truth LFO signal. 
The last two post-processing steps are only used for the unseen effect experiments in Section \ref{sec:unseen_fx}.

\subsection{Effect Modeling}\label{sec:effect_modeling}
\begin{figure}[t!]
    \centering
    \includegraphics[width=0.75\linewidth,trim={0.1cm 0.1cm 0.1cm 0.0cm},clip]{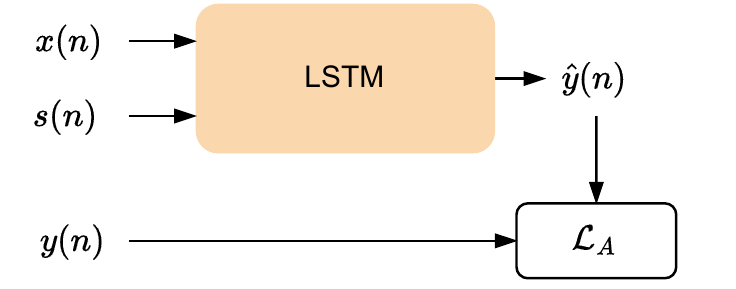}
    \vspace{-0.1cm}
    \caption{LFO effect modeling block diagram.}
    \label{fig:effect_model}
\end{figure}

Our effect model, shown in Figure~\ref{fig:effect_model}, is based on previous work in black-box modeling of modulation effects \cite{wright2021neural}. It consists of a long short-term memory (LSTM) network with a time-distributed linear layer that compresses the latent space into a single sample-by-sample value which is added to the input audio and then bounded by a hyperbolic tangent activation. The network takes as input two channels: the dry audio and an LFO conditioning signal. 

\linesubsec{Training}
Training is done on blocks of 1024 samples using truncated backpropagation through time (TBPTT) 
with 1024 samples of warmup. Once again, the AdamW
optimizer is used to minimize $\mathcal{L}_A$: the $L_1$ loss between the output audio and the ground truth wet audio. We do not train using Error-to-Signal Ratio (ESR) and DC loss as in \cite{wright2019real} since in our experiments we found using just the $L_1$ loss resulted in better results across all metrics.

\begin{table*}[t]
  \caption{Parameter values for the ``fixed params'' and ``varying params'' evaluation configurations.}
	\centering
	\begin{tabular}{l l c c c c c c c c c}
		\toprule
            && \multicolumn{3}{c}{LFO Parameters} & \multicolumn{6}{c}{Effect Parameters} \\  
            \cmidrule(lr){3-5} \cmidrule(lr){6-11}
		Effect  & Config.  & Shape & Phase & Rate & Center Freq. & Min. Delay & Delay Width         & Feedback  & Depth       & Mix \\ 
            \midrule
            Phaser  & Fixed   & Cos. & 0 - 2$\pi$ & 0.5 - 3.0\,Hz & 440\,Hz       & -        & -           & 0.25       & 1.0         & 1.0 \\
                    & Varying & & & & 70 - 18k\,Hz  & -        & -           & 0.0 - 0.7 & 0.25 - 1.0  & 1.0 \\
            \midrule 
            Flanger & Fixed   & All & 0 - 2$\pi$ & 0.5 - 3.0\,Hz & -          & 1 ms       & 4 ms        & 0.25       & 1.0         & 1.0 \\
                    & Varying & & & & -          & 0 - 1 ms   & 2.5 - 10 ms   & 0.0 - 0.7 & 0.25 - 1.0  & 1.0 \\
            \midrule 
            Chorus  & Fixed   & All & 0 - 2$\pi$ & 0.5 - 3.0\,Hz & -          & 20 ms    & 10 ms       & 0.25       & 1.0         & 1.0 \\
                    & Varying & & & & -          & 11 - 30 ms & 2.5 - 10 ms & 0.0 - 0.7 & 0.25 - 1.0  & 1.0 \\
            \bottomrule
	\end{tabular}
	\label{tab:params}
\end{table*}

\section{Experiments}

\subsection{Modulation Extraction}
Most phaser, flanger, and chorus implementations 
do not allow defining an arbitrary LFO signal. As a result, in order to be able to train the LFO extraction model with effects using arbitrary LFO signals, we implement our own flanger/chorus effect directly in PyTorch so that it can run on GPU and be integrated into our data pipelines. We use six different LFO shapes: cosine (cos), triangle (tri), rectified cosine (rect. cos), inverse rectified cosine (inv. rect. cos), sawtooth (saw), and inverse sawtooth (inv. saw). The LFO parameters of the module are phase, rate, and shape and the effect parameters are min. delay, delay width, feedback, depth, and mix. For the flanger effect we set the minimum delay to 0-1\,ms whereas for a chorus effect we set it to 10-20\,ms. We also use a modified version of the phaser provided in \texttt{Pedalboard}\footnote{\href{https://github.com/spotify/pedalboard}{https://github.com/spotify/pedalboard}}, which allows us to specify the LFO phase, while its shape remains restricted to a cosine waveform. Its LFO parameters are phase and rate and its effect parameters are center frequency, feedback, depth, and mix.

\linesubsec{Dataset}
We use the fourth subset of the IDMT-SMT-Guitar~\cite{kehling2014automatic} dataset, which contains 64 short electric guitar pieces grouped by genre. Each piece has been recorded at a fast and a slow tempo using three different guitars. We remove the two bars of synchronization tones at the beginning of each piece and split into 75\% training and 25\% validation sets across the 64 unique songs. This results in 154\,min of audio in the training set and 50\,min of audio in the validation set. We generate LFO signals with random phase, shape, and rate between 0.5 and 3\,Hz and then apply the three audio effects to random 2-second chunks of the dataset while uniformly sampling the effect parameters within their usable ranges.

\linesubsec{Training}
The input to LFO-net is a Mel spectrogram with 1024 FFT size, 256 sample hop length, 256 Mel bins, and a sample rate of 44.1\,kHz. The model consists of 6 convolutional blocks, each with 64 channels, a kernel size of 5 by 13, and a frequency max-pooling and temporal dilation factor of 2. As a result, the receptive field of the network along the time axis spans 2 seconds and outputs 345 frames given 88200 input samples. SpecAugment
of 25\% is applied during training to both the frequency and time axes. 
The model contains 1.3\,M parameters.

\linesubsec{Evaluation}
\label{sec:lfo_extraction_eval}
During evaluation of LFO-net, we smooth the signal using a 4th order moving average filter and keep phase, shape, and rate of the LFO signal random. We define two different effect parameter configurations to compare against: ``fixed params'' and ``varying params'', summarized in Table \ref{tab:params}. We evaluate on 1000 random 2-second non-silent chunks of the dataset.
As a baseline, we assume an experienced audio engineer could correctly guess the shape of the LFO signal, whether it’s going up or down, and the approximate rate of modulation from listening to the wet audio. We define this as an LFO signal with the correct shape, a random phase error of up to 50\%, and a random rate error of up to 25\%. 

\subsection{Unseen Audio Sources}
\label{sec:unseen_audio}

We evaluate the LFO-net on five unseen datasets processed with the \texttt{Pedalboard} phaser and our flanger/chorus implementation using the same setup described in the previous experiment (Section \ref{sec:lfo_extraction_eval}). These datasets are guitar, bass/double bass, and keyboard audio from MedleyDB 2.0~\cite{bittner2016medleydb}, drums from the IDMT-SMT-Drums~\cite{dittmar2014real} dataset, and vocals from VocalSet~\cite{wilkins2018vocalset}.

\subsection{Quasiperiodic, Combined, and Distorted Modulations}
\begin{figure}[t]
\centering
\begin{minipage}{.32\linewidth}
    \centering
    \includegraphics[width=1.0\linewidth]{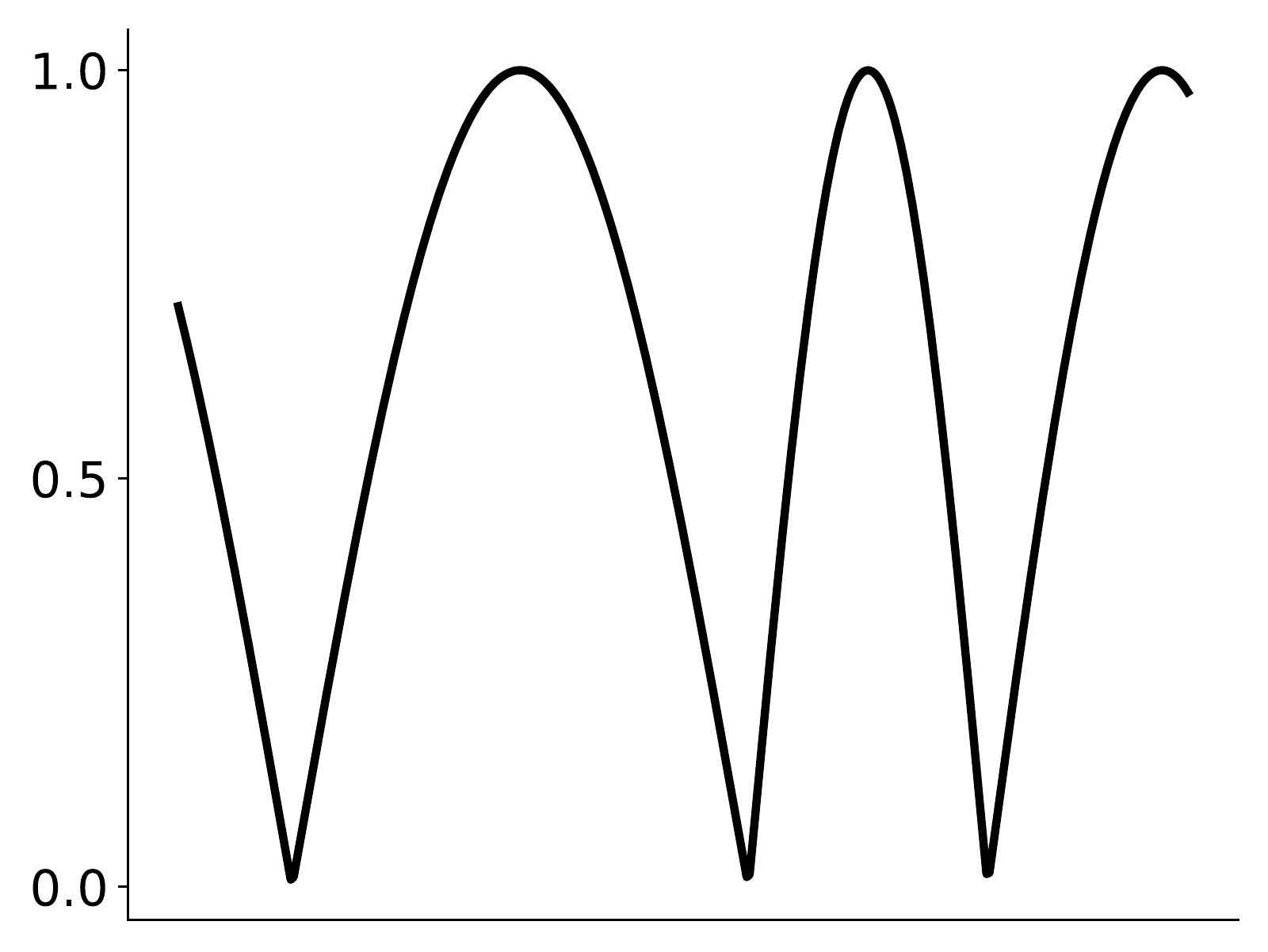}
\end{minipage}
\begin{minipage}{.32\linewidth}
    \centering
    \includegraphics[width=1.0\linewidth]{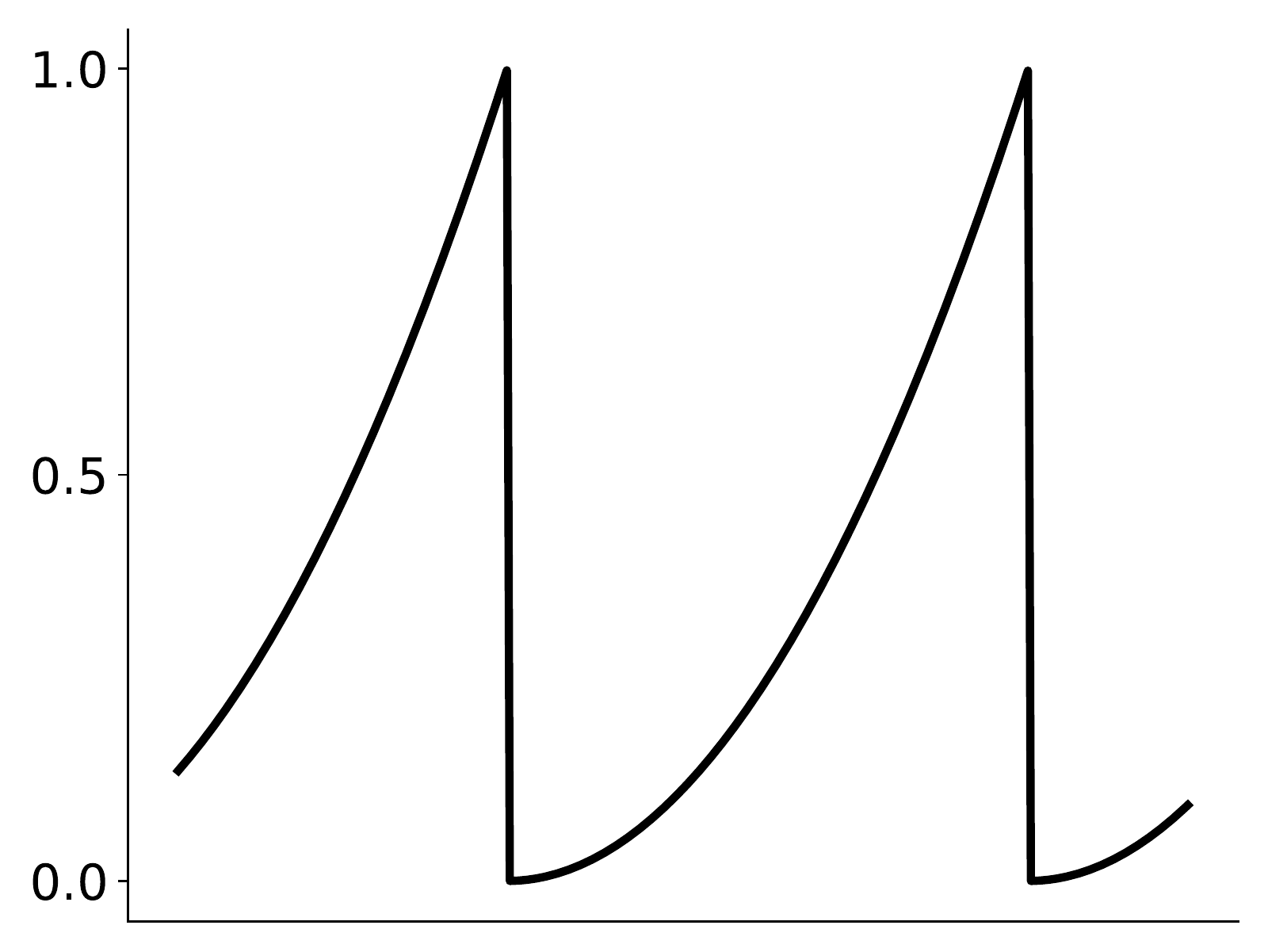}
\end{minipage}
\begin{minipage}{.32\linewidth}
    \centering
    \includegraphics[width=1.0\linewidth]{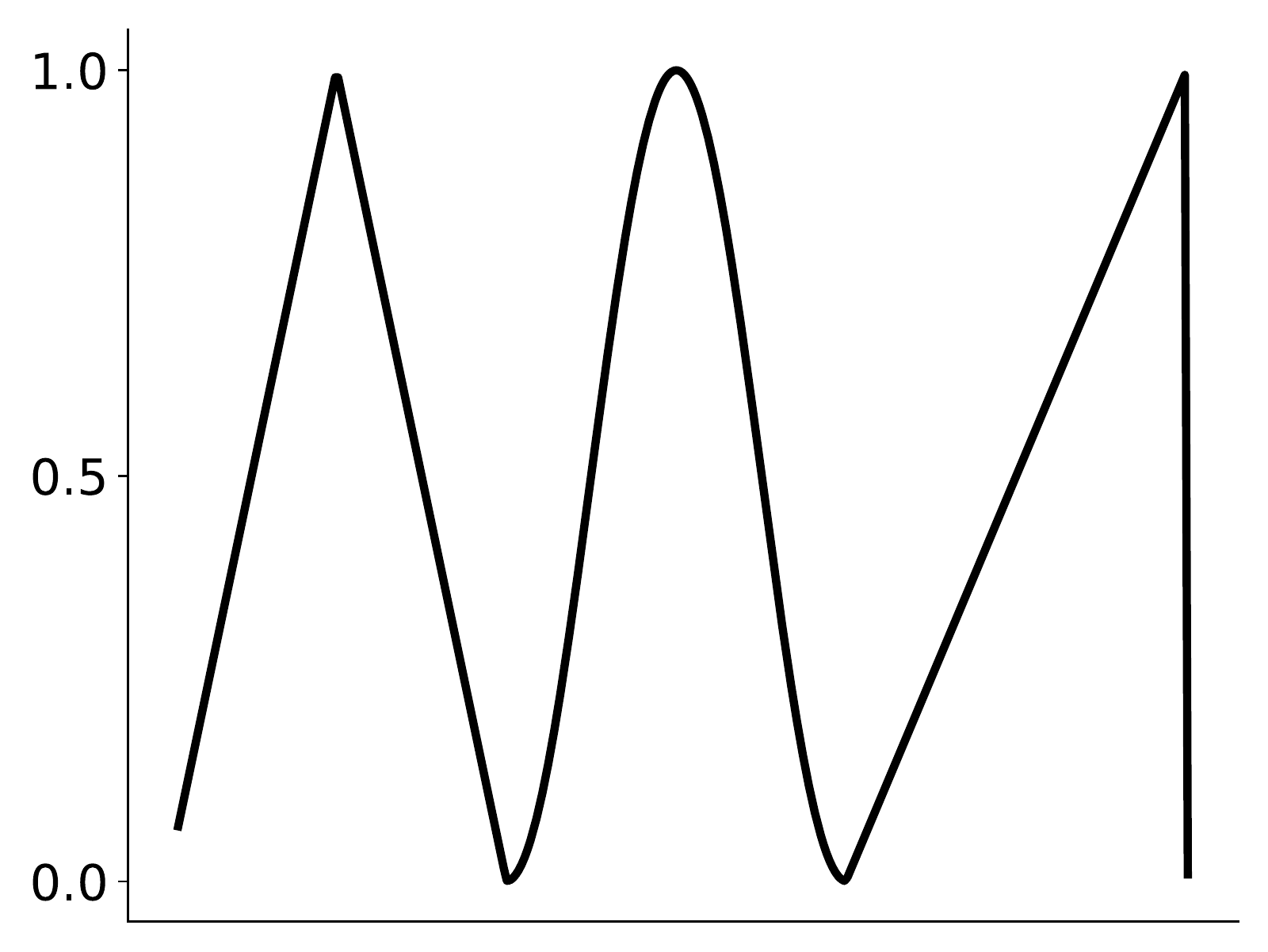}
\end{minipage}
\caption{Examples of quasiperiodic (left), distorted (center), and combined (right) LFO shapes.}
\label{fig:lfo_irregular_examples}
\vspace{-0.0cm}
\end{figure}

Irregular LFO shapes can greatly expand the creative possibilities of an effect and are commonplace in virtual synthesizers. Furthermore, the internal LFOs of analog audio effects are imperfect and can drift or become distorted. As a result, we test the ability of our LFO extraction model to generalize to irregular LFO shapes. 
We generate quasiperiodic LFO signals by randomly stretching each cycle of a periodic modulation by 10 - 33.33\%. We generate combined LFO signals by swapping out random cycles of a periodic modulation with a different shape. We try combining all six shapes randomly together and the four symmetrical shapes (no sawtooth and inverse sawtooth). Finally, we distort LFO signals via exponentiation, which makes different sections of the signal more concave or convex. 
Figure \ref{fig:lfo_irregular_examples} shows examples of these three types of irregular LFO signals. 
We then evaluate on the test dataset using the ``fixed params'' evaluation configuration.

\subsection{Latent Space Visualization}
In order to see whether the model learns meaningful representations in its latent space we generate three different visualizations. 
We perform inference on 200 samples from the validation dataset while changing one variable and keeping all others fixed. A single 64-dimensional latent vector is obtained by taking the average across the output frames of the final convolutional block in the model. 
We then produce a 2d visualization of the vectors using principle component analysis (PCA).
We explore how the rate and shape of the LFO signal are encoded as well as the difference between the phaser, flanger, and chorus effects.

\begin{figure}[t!]
\centering
\begin{minipage}{.32\linewidth}
    \centering
    \includegraphics[width=1.0\linewidth]{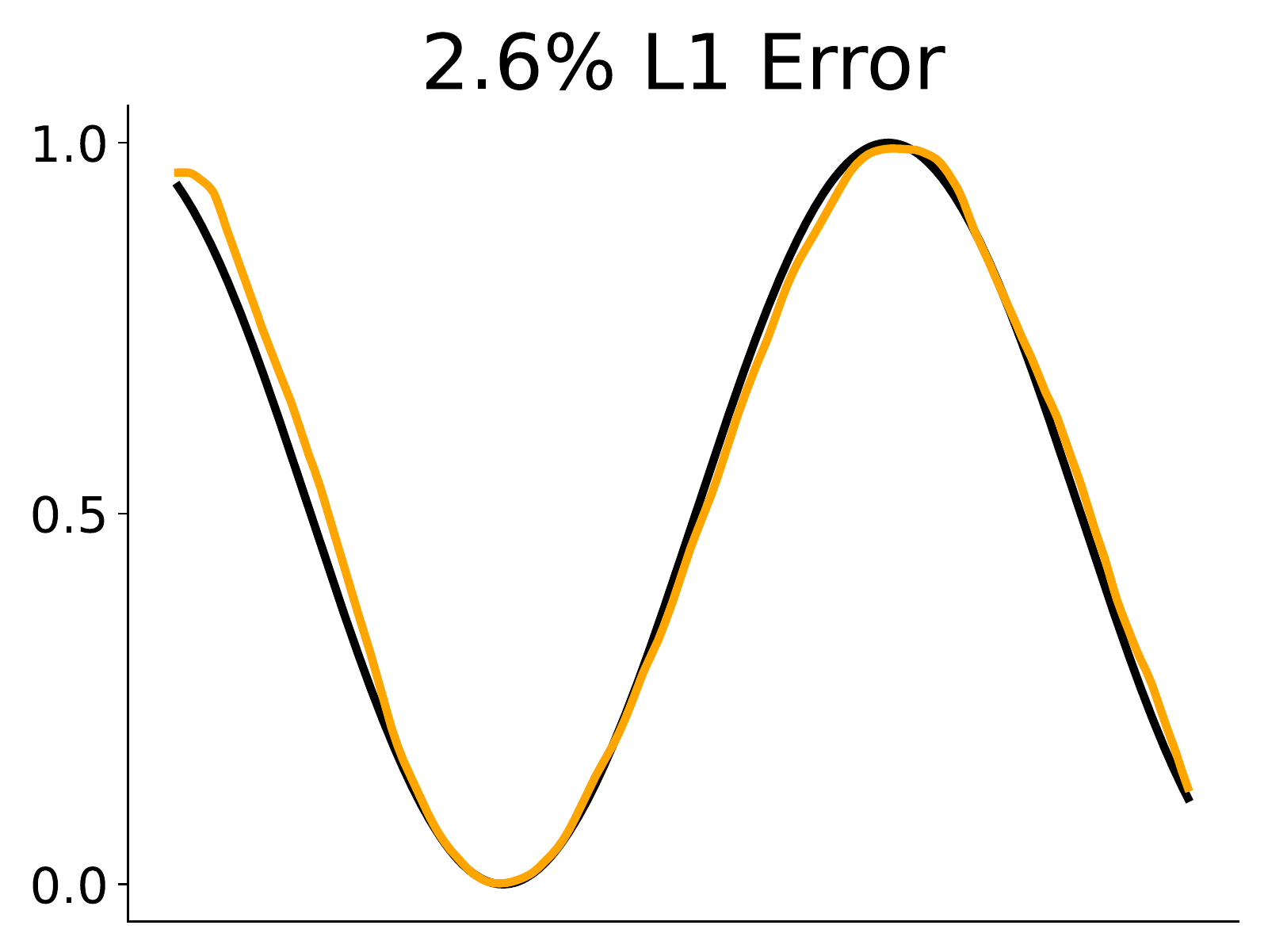}
\end{minipage}
\begin{minipage}{.32\linewidth}
    \centering
    \includegraphics[width=1.0\linewidth]{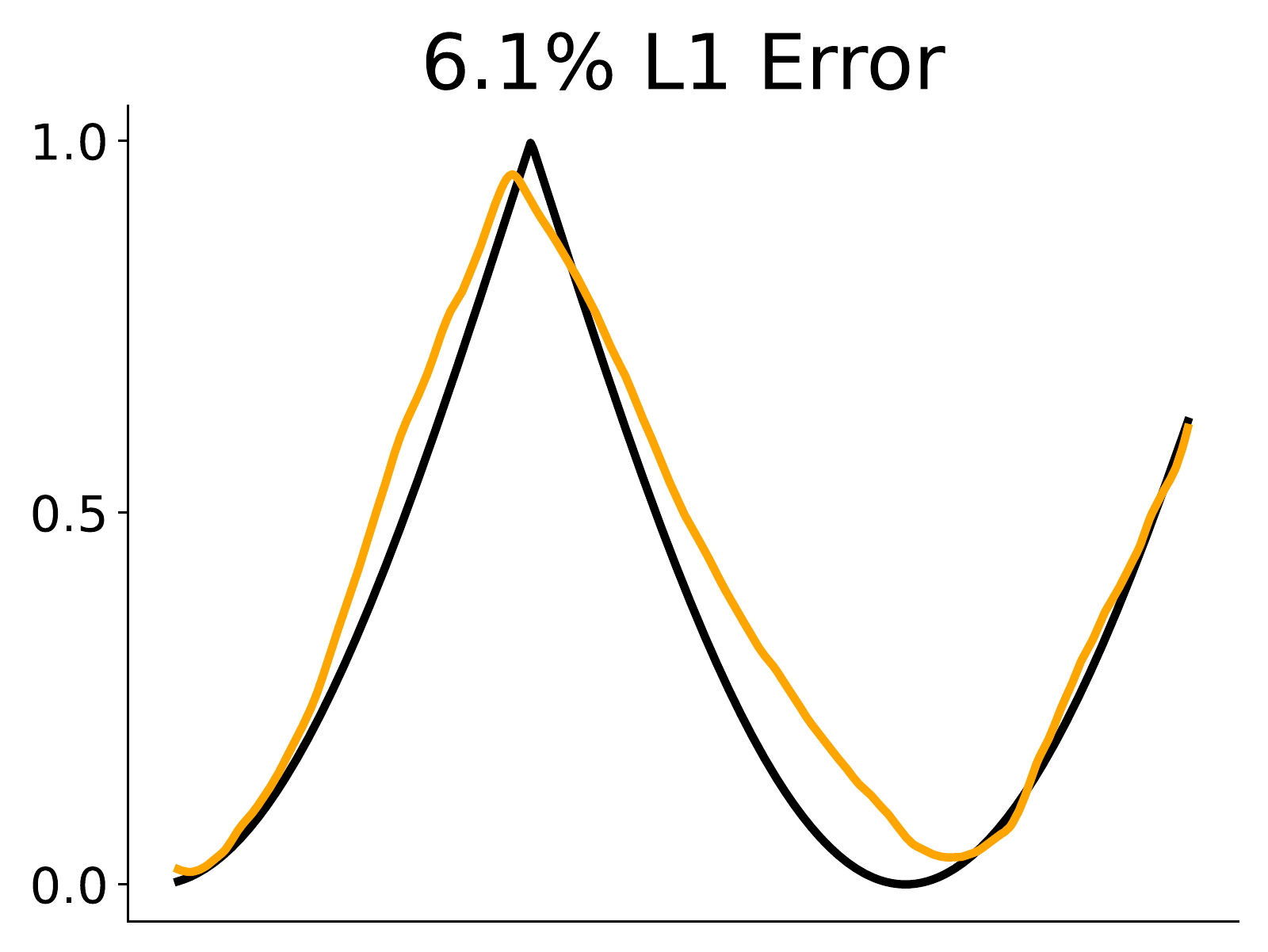}
\end{minipage}
\begin{minipage}{.32\linewidth}
    \centering
    \includegraphics[width=1.0\linewidth]{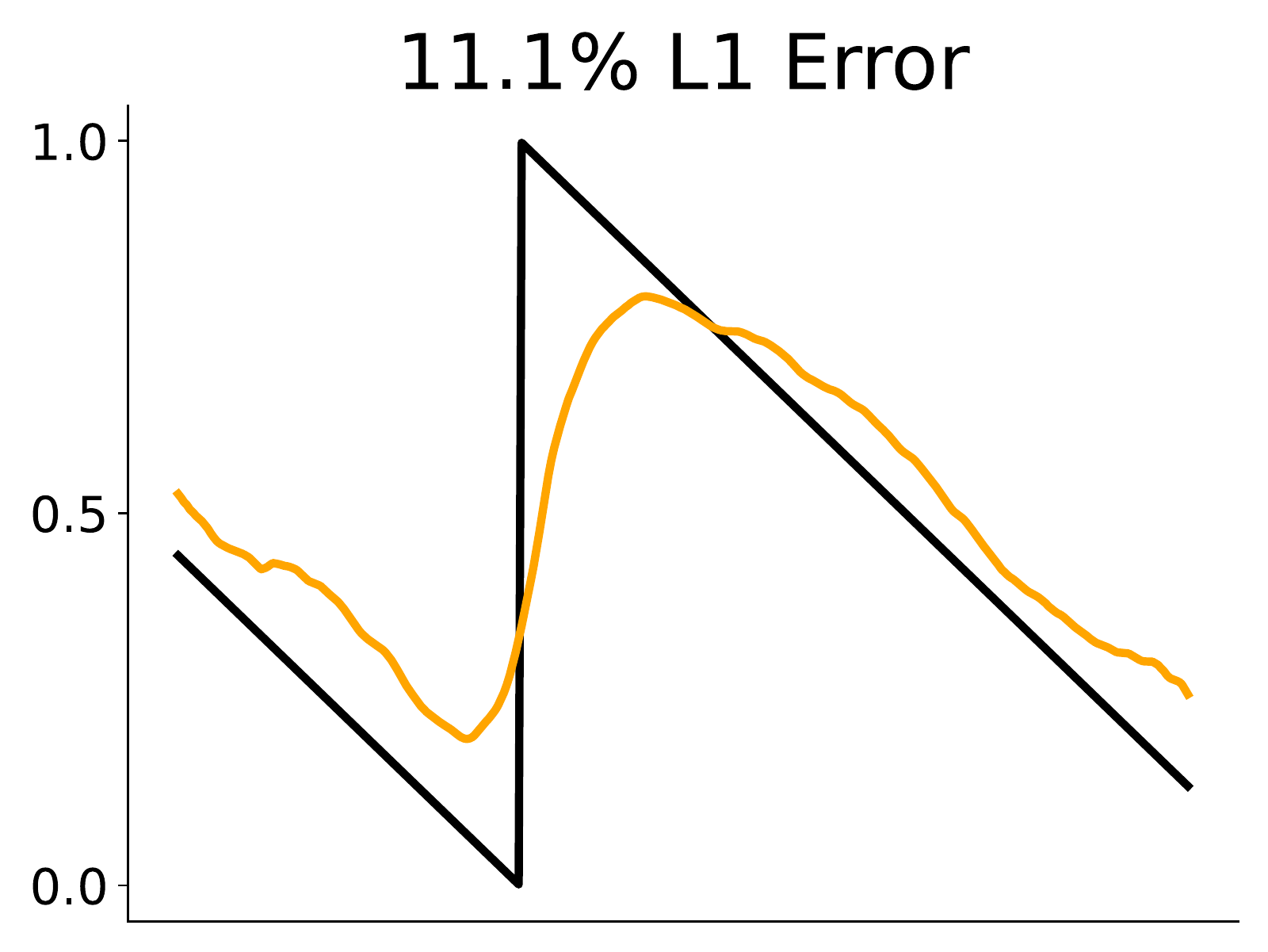}
\end{minipage}
\caption{Examples of 3\%, 6\%, and 11\% extracted LFO $L_1$ errors.}
\label{fig:l1_errors}
\end{figure}

\subsection{Unseen Analog and Digital Effects}
\label{sec:unseen_fx}

Our final experiment evaluates whether LFO-net can be applied to unseen analog and digital effects and then be used to condition and train an effect emulation model. 
We use the EGFxSet dataset~\cite{pedroza2022egfxset}, which consists of five-second long recordings of single electric guitar notes processed with an MXR Phase 45 phaser pedal, a Mooer E-Lady flanger pedal, and a Boss CE-3 chorus pedal. 
Referencing the datasheets of these effects, we established that all three effects use a rounded triangle LFO shape. 
We peak normalize the input chunks of audio since the volume levels differ significantly between the wet and dry audio pairs. 
We apply all post-processing steps described in Section~\ref{sec:postprocessing} during training and inference when extracting the LFO signal (8th order moving average for smoothing) and use a 70/18/12\% train-val-test split.

\begin{figure}[t]
    \centerline{\includegraphics[width=0.85\linewidth,clip]{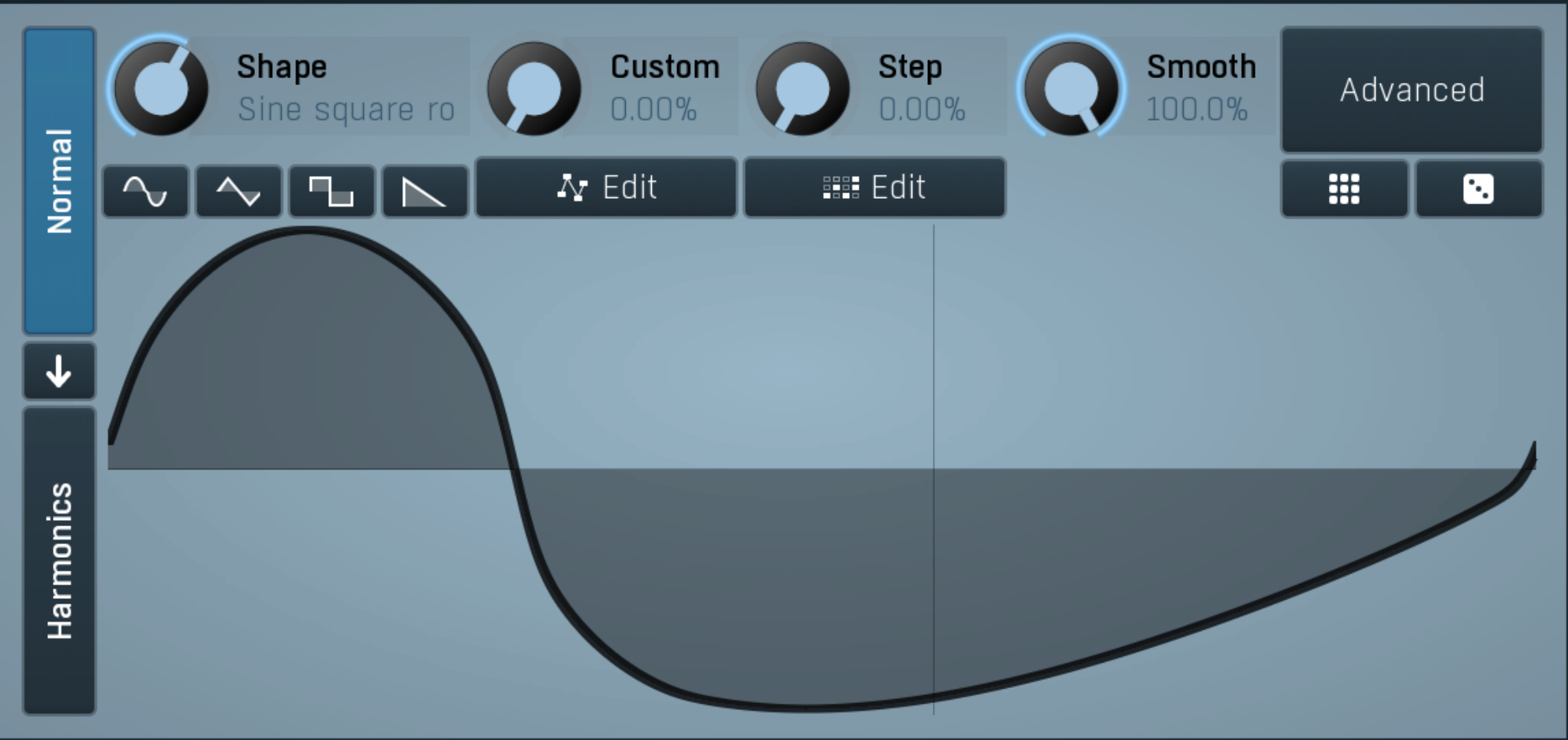}}
    \caption{Skewed sinusoidal LFO shape used in the Melda Phaser and Flanger irregular LFO effect modeling experiments.}
    \label{fig:melda_lfo}
    \vspace{-0.2cm}
\end{figure}

For digital effects we use the MeldaProduction MPhaser and MFlanger plugins\footnote{\href{https://www.meldaproduction.com/effects/free}{https://www.meldaproduction.com/effects/free}}.
These effects give the user control over the LFO signal and enable combined and irregular LFO signals to be drawn in the user interface. 
We test two scenarios. 
First, we consider modeling a phaser and flanger effect with an irregular LFO signal and then with a quasiperiodic LFO signal. 
For the irregular case we define a skewed sinusoidal LFO shape as shown in Figure \ref{fig:melda_lfo} at a frequency of 0.75\,Hz. 
For the quasiperiodic case we start with a triangle shape and automate the rate of the LFO from 0.5\,Hz to 2.0\,Hz and back every 4 seconds. 
We apply both effects to 8 minute training, 2.5 minute validation, and 2 minute test sets from the fourth subset of the IDMT-SMT-Guitar dataset. 
During post-processing for the irregular case, we omit step 2 (stretching) to preserve the original shape of the extracted LFO signal.

Since we do not have access to the internal LFO signal for these effects, we first confirm visually that the LFO extraction model is able to output similar LFO signals when applied to these unseen effects. We then use it to train one effect model LSTM with 64 hidden units for each of the seven analog and digital effect configurations defined previously that learns to reconstruct the wet audio given the dry audio and the extracted LFO signal. As a baseline we also train effect models conditioned on a randomly generated LFO with a triangle shape and 0-25\% frequency error for the analog effects, a triangle shape and random frequency between 0.5 - 2.0\,Hz for the quasiperiodic experiment, and a cosine shape and 0-25\% frequency error for the irregular LFO signal experiment.

\vspace{-4pt}
\section{Results}
\vspace{-4pt}

\subsection{Modulation Extraction} Table \ref{tab:eval_val} summarizes the ability of the model to extract LFO signals from the test dataset. 
Figure \ref{fig:l1_errors} provides a visual reference for the reconstruction quality corresponding to different $L_1$ error values. 
We find that an $L_1$ error of less than 5\% corresponds to very accurate extraction with less than 10\% error still being acceptable. We notice that the model struggles most with the asymmetrical sawtooth shapes. This is likely due to the waveform containing sharp edges, which can be difficult to reconstruct. We also observe that the model is better at extracting the LFO from the phaser, and worse at extracting the LFO from the chorus. This matches our intuition since the phaser is limited to a cosine LFO shape and because the chorus effect contains the largest varying delay which results in the greatest change in the wet audio spectrogram compared to the flanger. Finally, there is no difference in model performance when the parameters are fixed or varying across their entire usable ranges, thus highlighting the learning capabilities of the proposed LFO model architecture. The baseline consistently results in very large errors due to the fact that small differences in phase and frequency can cause the baseline and ground truth signal to drift apart. We also experimented with extracting the LFO signal from just the wet audio (no dry audio channel) and found that this resulted in an approximately 3\% increase in the $L_1$ error.

\begin{table}[t]
    \caption{LFO extraction evaluation metrics.}
	\centering
	\begin{tabular}{ll c c c} \toprule
            && \multicolumn{3}{c}{$L_1$ Error (\%)} \\ \cmidrule(lr){3-5}
		Effect & LFO Shape & Fixed &  Varying & Baseline \\ \midrule
            Phaser  & Cosine            & 1.8\% & 2.1\% & 32\% \\ \midrule
            Flanger & Cosine            & 1.9\% & 1.9\% & 32\% \\ 
                    & Triangle          & 2.2\% & 2.3\% & 27\% \\
                    & Rect. Cosine      & 2.2\% & 2.1\% & 28\% \\
                    & Inv. Rect. Cos.   & 1.9\% & 2.0\% & 28\% \\
                    & Saw               & 4.5\% & 4.5\% & 27\% \\
                    & Inv. Saw          & 4.9\% & 4.7\% & 27\% \\
                    & All               & 2.9\% & 2.9\% & 28\% \\
            \midrule
            Chorus  & Cosine            & 3.6\% & 2.9\% & 32\% \\
                    & Triangle          & 3.1\% & 3.3\% & 27\% \\
                    & Rect. Cosine      & 2.7\% & 2.9\% & 28\% \\
                    & Inv. Rect. Cos.   & 2.9\% & 2.9\% & 28\% \\
                    & Saw               & 8.0\% & 6.9\% & 27\% \\
                    & Inv. Saw          & 8.5\% & 7.3\% & 27\% \\
                    & All               & 4.7\% & 4.3\% & 28\% \\
            \midrule
            All     & All               & 3.1\% & 3.1\% & 29\% \\
            \bottomrule
	\end{tabular}
	\label{tab:eval_val}
\end{table}

\subsection{Unseen Audio Sources} We find the model generalizes well to unseen data processed with our three training effects. From Table \ref{tab:eval_unseen_data} we see that LFO-net performs just as well or even better on the unseen guitar, bass, and keys datasets. Performance on vocals is also only marginally worse. We expect drums to be the most challenging to extract LFO signals from due to the less tonal and dense onsets and the results match this intuition with extraction ability becoming worse for the flanger and chorus effects on the drums dataset. Varying parameters also results in a very small reduction in performance compared to fixed parameters.

\setlength{\tabcolsep}{4.0pt}
\begin{table}[t!]
  \caption{LFO extraction metrics for unseen datasets.}
	\centering
	\begin{tabular}{llrrrr} \toprule
            && \multicolumn{4}{c}{$L_1$ Error (\%)} \\ \cmidrule(lr){3-6}
		Dataset    & Params & Phaser & Flanger & Chorus & All \\ \midrule
            MDB Guitar & Fixed   & 1.8\%  & 2.8\%   & 4.7\%  & 3.1\% \\
                       & Varying & 1.8\%  & 2.8\%   & 4.9\%  & 3.2\% \\  \midrule
            MDB Bass   & Fixed   & 1.9\%  & 2.4\%   & 4.3\%  & 2.9\% \\
                       & Varying & 2.3\%  & 2.6\%   & 4.7\%  & 3.2\% \\  \midrule
            MDB Keys   & Fixed   & 1.8\%  & 2.5\%   & 4.2\%  & 2.8\% \\
                       & Varying & 2.3\%  & 2.5\%   & 4.0\%  & 2.9\% \\  \midrule
            IDMT Drums & Fixed   & 1.9\%  & 5.3\%   & 12.2\%  & 6.5\% \\
                       & Varying & 2.7\%  & 5.8\%   & 11.3\%  & 6.6\% \\  \midrule
            Vocalset   & Fixed   & 2.8\%  & 4.3\%   & 5.4\%  & 4.2\% \\
                       & Varying & 2.7\%  & 4.2\%   & 5.8\%  & 4.2\% \\ \bottomrule   
	\end{tabular}
	\label{tab:eval_unseen_data}
\end{table}

\subsection{Quasiperiodic, Combined, and Distorted}

The quasiperiodic, distorted, and combined LFO signal results are contained in Tables \ref{tab:eval_quasi} and \ref{tab:eval_combination}. The ability to extract quasiperiodic signals is only slightly reduced when compared to periodic signals with the chorus and asymmetrical shapes appearing more challenging than the flanger and symmetrical shapes. This implies the system could be used to obtain an LFO signal for non-periodic audio effects. 

Distorted inverse rectified cosine, saw, and inverse saw are also difficult for LFO-net to extract. We believe this is because the inverse rectified cosine shape becomes closer to a square wave at the troughs when exponentiated which results in a constant delay and less sweeping patterns in the spectrum to analyze. Similarly, the saw and inverse saw shapes become even more jagged at the corners, thus making reconstruction more challenging, especially at higher LFO rates.
Finally, we found that LFO-net is better at reconstructing random combinations of the LFO shapes when the asymmetrical ones are omitted. We believe this is due to the harsh discontinuities that can be introduced by combining sawtooth waves with the other symmetrical waves. Our results indicate that the model can extract symmetrical modulation shapes well, even when each period consists of a different shape.

\begin{table}[t!]
  \caption{LFO extraction metrics for quasi. and distorted signals.}
	\centering
	\begin{tabular}{l l c c c c}
		\toprule
            && \multicolumn{4}{c}{$L_1$ Error (\%)} \\ \cmidrule(lr){3-6}
		Effect & LFO Shape & Quasi. & Base. & Dist. & Base. \\ \midrule
            Flanger & Cosine            & 3.3\% & 32\% & 3.4\% & 33\% \\
                    & Triangle          & 3.6\% & 28\% & 2.4\% & 30\% \\
                    & Rect. Cosine      & 3.7\% & 28\% & 1.9\% & 32\% \\
                    & Inv. Rect. Cos.   & 3.3\% & 29\% & 8.1\% & 28\% \\
                    & Saw               & 5.8\% & 27\% & 13\%  & 32\% \\
                    & Inv. Saw          & 6.5\% & 28\% & 13\%  & 31\% \\
                    & All               & 4.5\% & 29\% & 6.7\% & 31\% \\
             \midrule
            Chorus  & Cosine            & 4.7\% & 32\% & 4.6\% & 33\% \\
                    & Triangle          & 5.3\% & 28\% & 3.1\% & 30\% \\
                    & Rect. Cosine      & 4.9\% & 28\% & 3.6\% & 32\% \\
                    & Inv. Rect. Cos.   & 4.3\% & 29\% & 8.7\% & 28\% \\
                    & Saw               & 10\%  & 27\% & 16\%  & 32\% \\
                    & Inv. Saw          & 11\%  & 28\% & 16\%  & 31\% \\
                    & All               & 7.0\% & 29\% & 8.5\% & 31\% \\
             \midrule
            Both    & All               & 5.8\% & 29\% & 7.6\% & 31\% \\
            \bottomrule
	\end{tabular}
	\label{tab:eval_quasi}
\end{table}

\begin{table}[t!]
  \caption{LFO extraction metrics for combined modulations.}
	\centering
	\begin{tabular}{l l c c}
		\toprule
            && \multicolumn{2}{c}{$L_1$ Error (\%)} \\ \cmidrule(lr){3-4}
		Effect & LFO Shapes & Combined & Baseline \\ 
            \midrule
            Flanger & Symmetrical & 4.7\% & 33\% \\
                    & All                                   & 9.4\% & 34\% \\
            \midrule
            Chorus  & Symmetrical & 6.1\% & 33\% \\
                    & All                                   & 11.2\% & 34\% \\
            \midrule
            Both    & Symmetrical & 5.4\% & 33\% \\
                    & All                                   & 10.3\% & 34\% \\
            \bottomrule
	\end{tabular}
	\label{tab:eval_combination}
\end{table}

\subsection{Latent Space Visualization} 
\begin{figure*}
\centering
\begin{minipage}{.31\textwidth}
  \centering
  \includegraphics[width=0.85\linewidth,trim={0.0cm 0.0cm 0.0cm 0},clip]{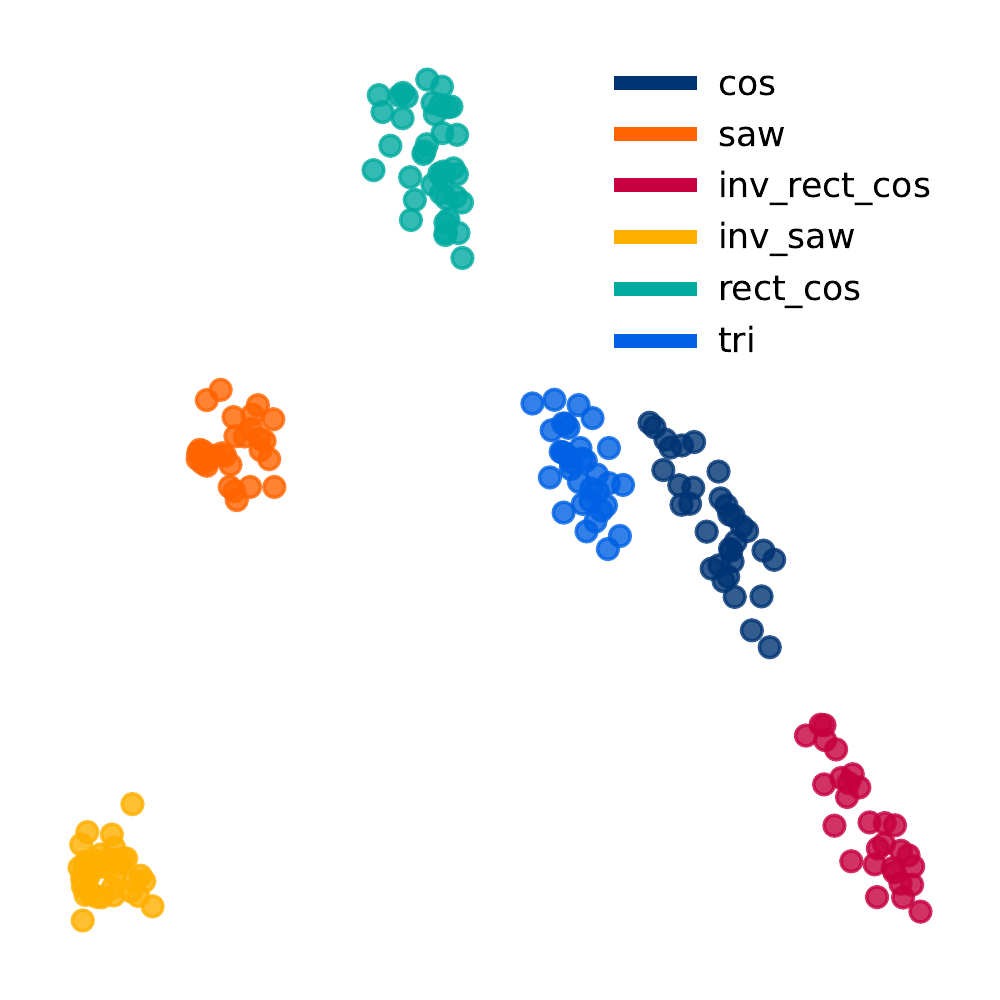}
  \captionof{figure}{LFO shape}
  \label{fig:latent_shape}
\end{minipage}%
\begin{minipage}{.31\textwidth}
  \centering
  \includegraphics[width=0.85\linewidth,trim={0.0cm 0.0cm 0.0cm 0},clip]{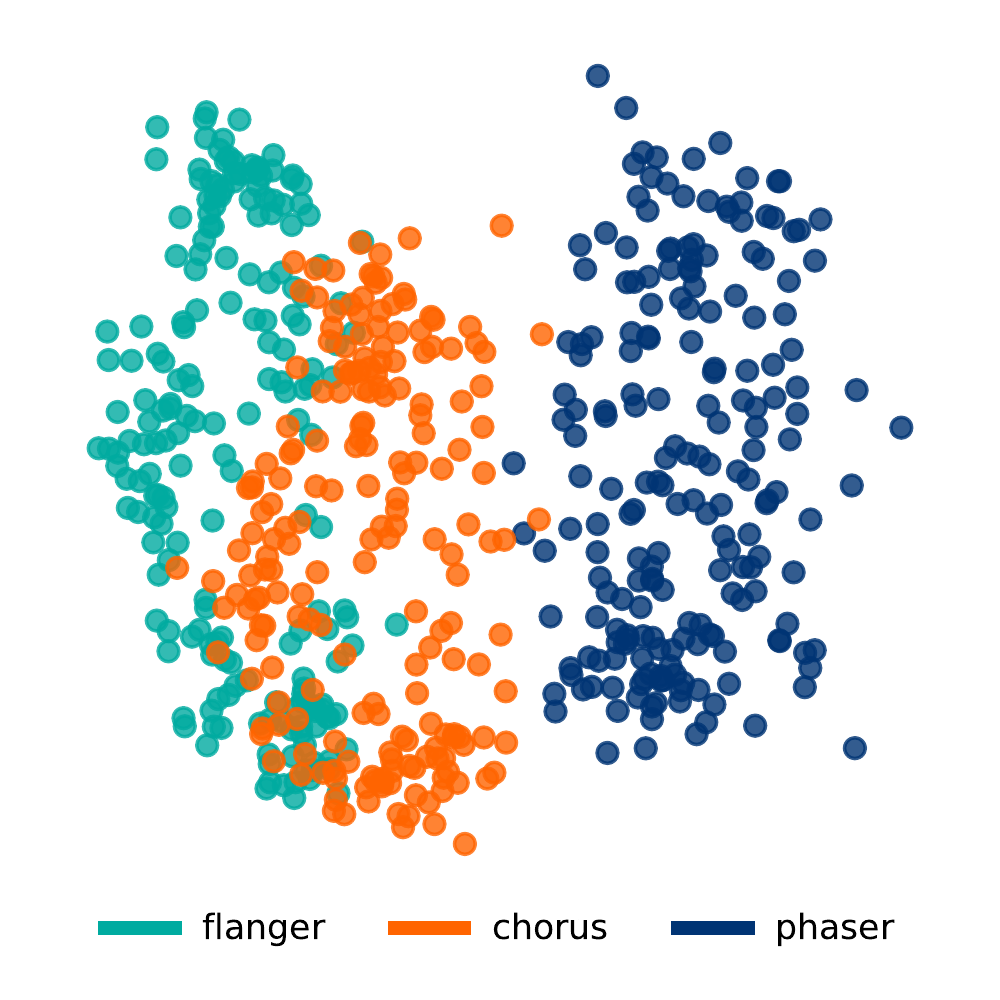}
  \captionof{figure}{LFO effect type}
  \label{fig:latent_effect}
\end{minipage}
\begin{minipage}{.33\textwidth}
  \centering
  \includegraphics[width=0.99\linewidth,trim={0.0cm 0.0cm 0.0cm 0},clip]{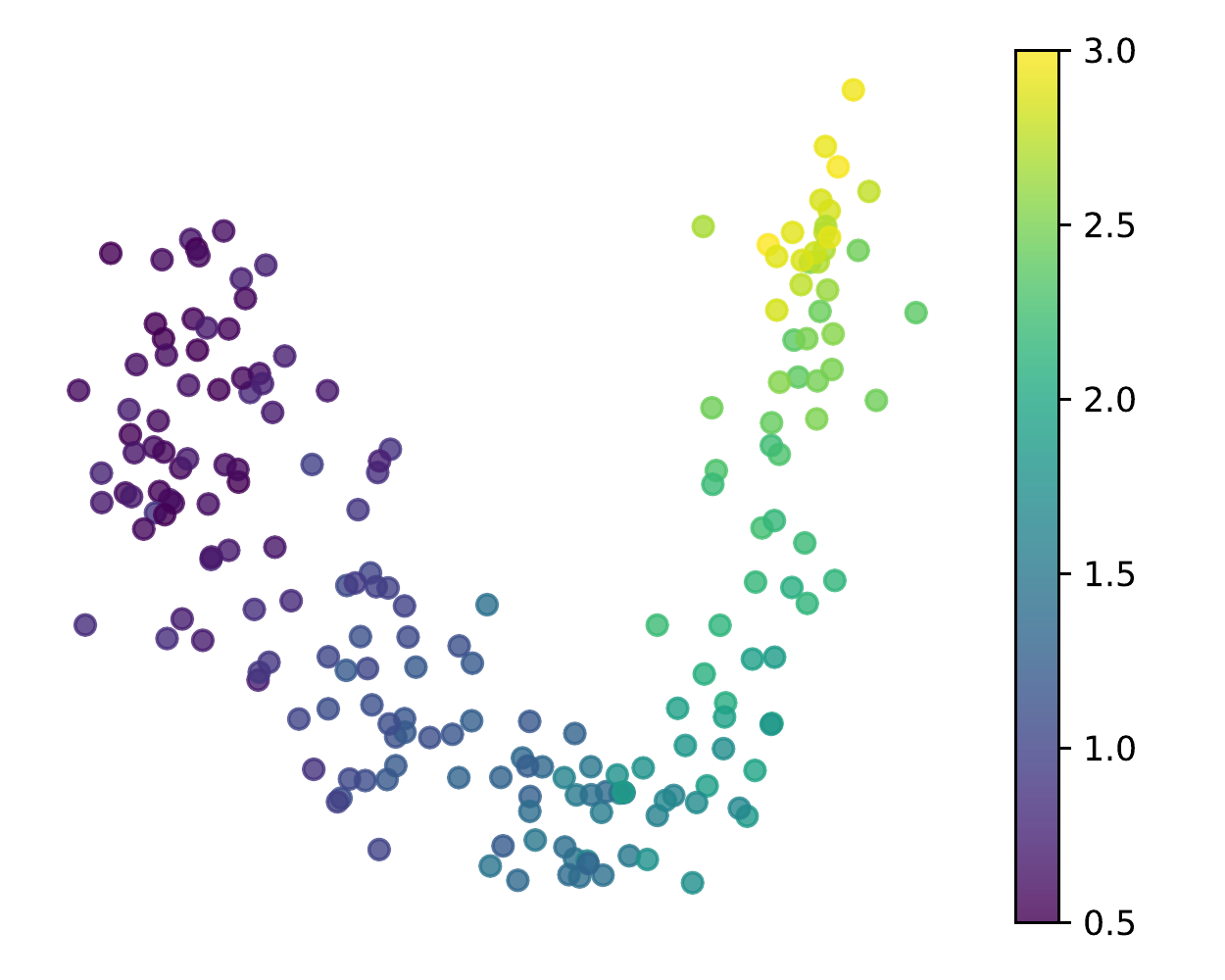}
  \captionof{figure}{LFO rate}
  \label{fig:latent_rate}
\end{minipage}
\end{figure*}

The latent space visualizations for changing LFO shape, effect, and rate are shown in Figures \ref{fig:latent_shape}, \ref{fig:latent_effect}, and \ref{fig:latent_rate}, respectively. The latent space decouples for all three visualizations with the relationship between different LFO shapes being encoded by the distance of their clusters in the latent space. Opposite pairs of shapes (i.e. saw / inverse saw and rect. cos. / inv. rect. cos.) are separated by a large distance and similar shapes like triangle and cosine are close together. Similarly, the three different effects decouple in the latent space with chorus and flanger having more overlap since they are identical in implementation, but with different delay amounts. We expect the phaser effect to be the most distinct since it is a unique implementation. Finally, the LFO rate visualization displays a clear relationship between the frequency of the LFO and position in the latent space, with high frequencies becoming more densely clustered together.

\begin{table}[t]
  \caption{Unseen effect evaluation results.}
	\centering
	\begin{tabular}{l l c c c c}
		\toprule
            && \multicolumn{2}{c}{Audio Error} & \multicolumn{2}{c}{Baseline Error} \\
            \cmidrule(lr){3-4} \cmidrule(lr){5-6}
		Effect & LFO Shape & $L_1$ (\%) & ESR & $L_1$ (\%) & ESR  \\ 
            \midrule
            EGFx Phaser   & Analog Tri. & 3.5\% & 0.42 & 6.1\% & 0.78 \\
            EGFx Flanger  & Analog Tri. & 5.8\% & 0.94 & 5.9\% & 0.95 \\
            EGFx Chorus   & Analog Tri. & 5.0\% & 0.59 & 6.6\% & 0.82 \\
            \midrule
            Melda Phaser  & Quasi. Tri.       & 1.4\% & 0.21 & 2.7\% & 0.61 \\
                          & Irregular    & 0.76\% & 0.08 & 3.0\% & 0.78 \\
            \midrule
            Melda Flanger & Quasi. Tri.       & 2.3\% & 0.13 & 5.3\% & 0.51 \\
                          & Irregular    & 2.9\% & 0.18 & 5.2\% & 0.45 \\
            \bottomrule
	\end{tabular}
	\label{tab:eval_em}
        \vspace{-0.4cm}
\end{table}

\subsection{Unseen Analog and Digital Effects} 
\begin{figure}[t]
\centering
\begin{minipage}{.32\linewidth}
    \centering
    \includegraphics[width=1.0\linewidth]{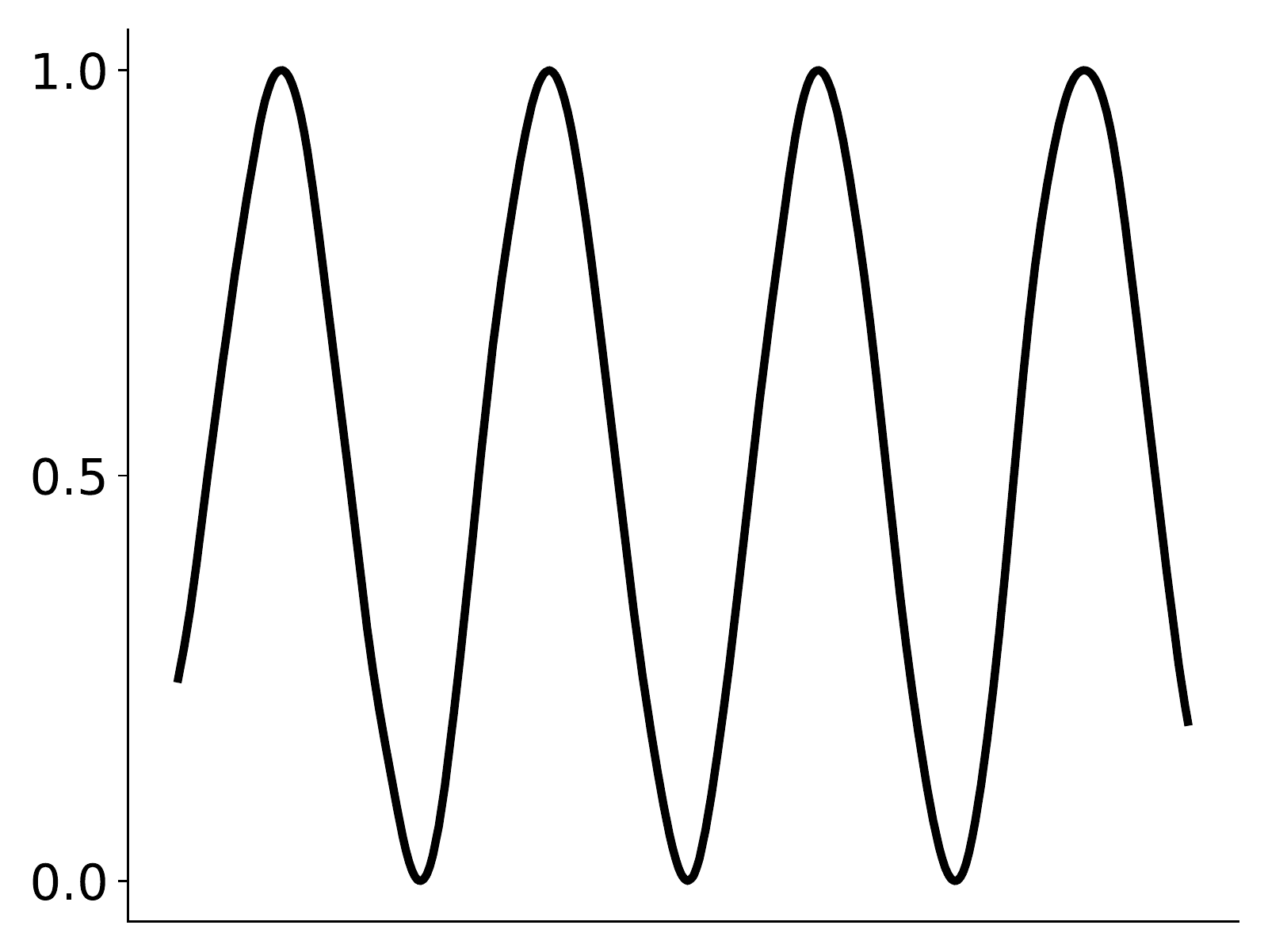}
\end{minipage}
\begin{minipage}{.32\linewidth}
    \centering
    \includegraphics[width=1.0\linewidth]{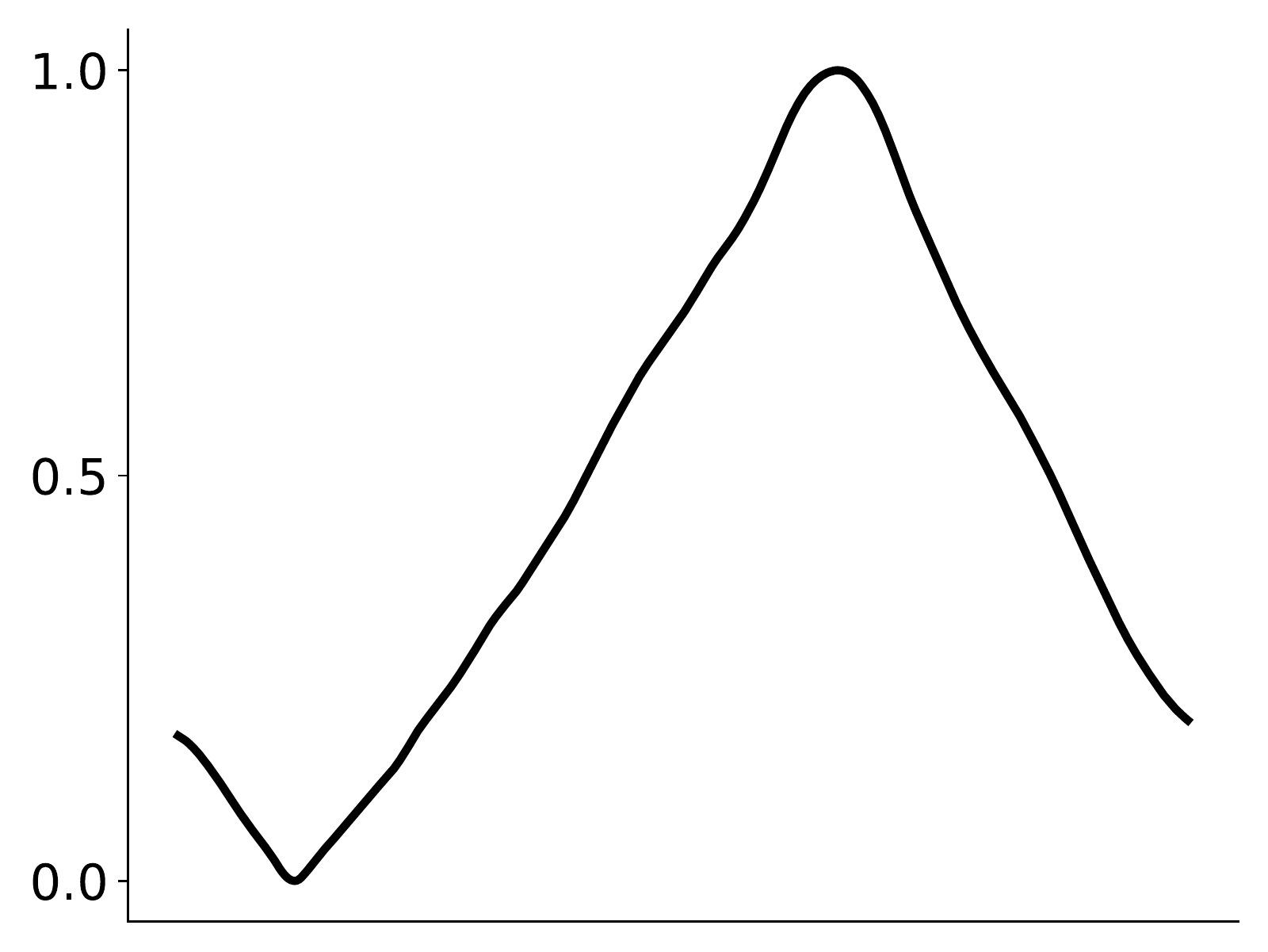}
\end{minipage}
\begin{minipage}{.32\linewidth}
    \centering
    \includegraphics[width=1.0\linewidth]{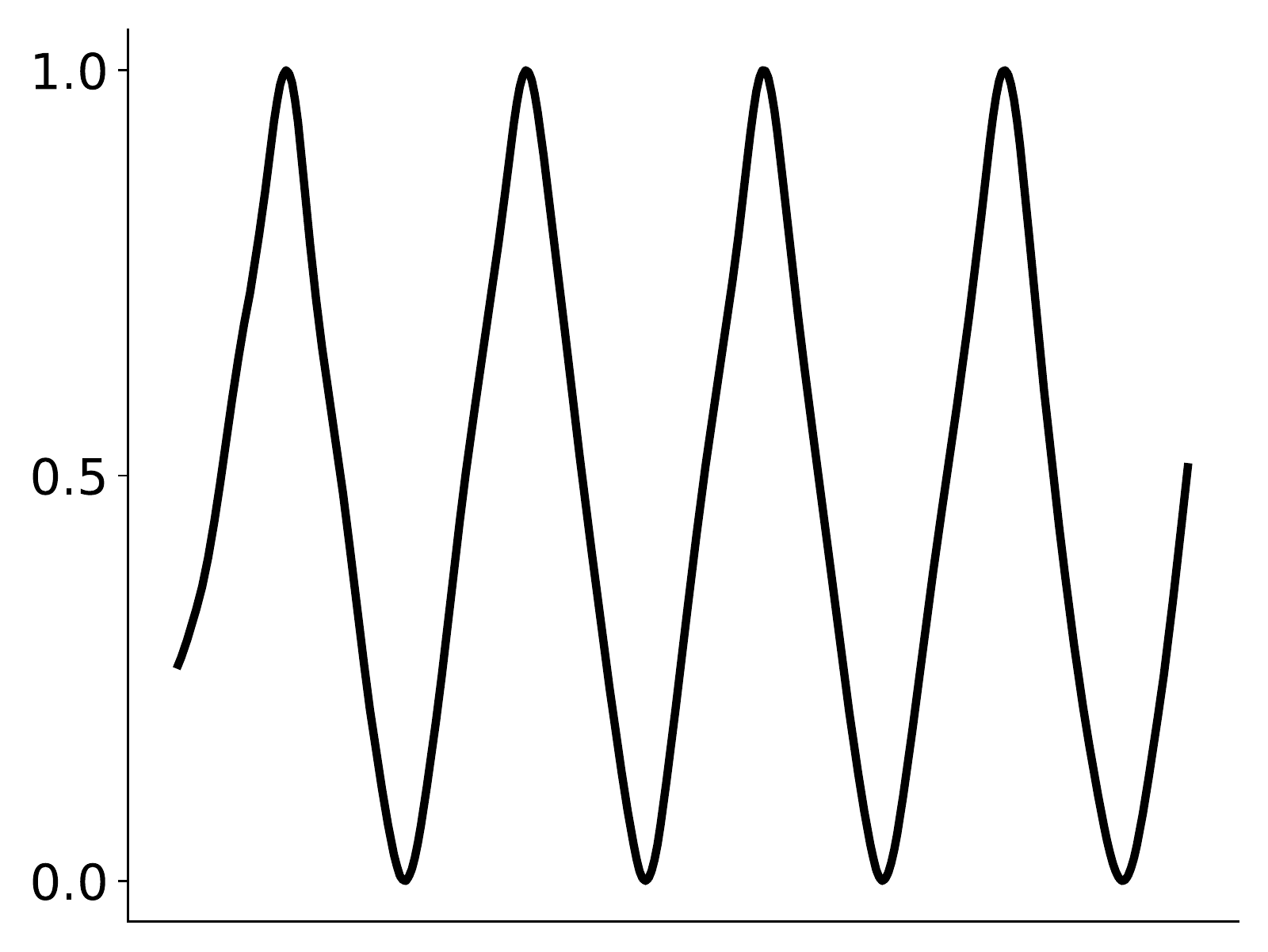}
\end{minipage}
\begin{minipage}{.32\linewidth}
    \centering
    \includegraphics[width=1.0\linewidth]{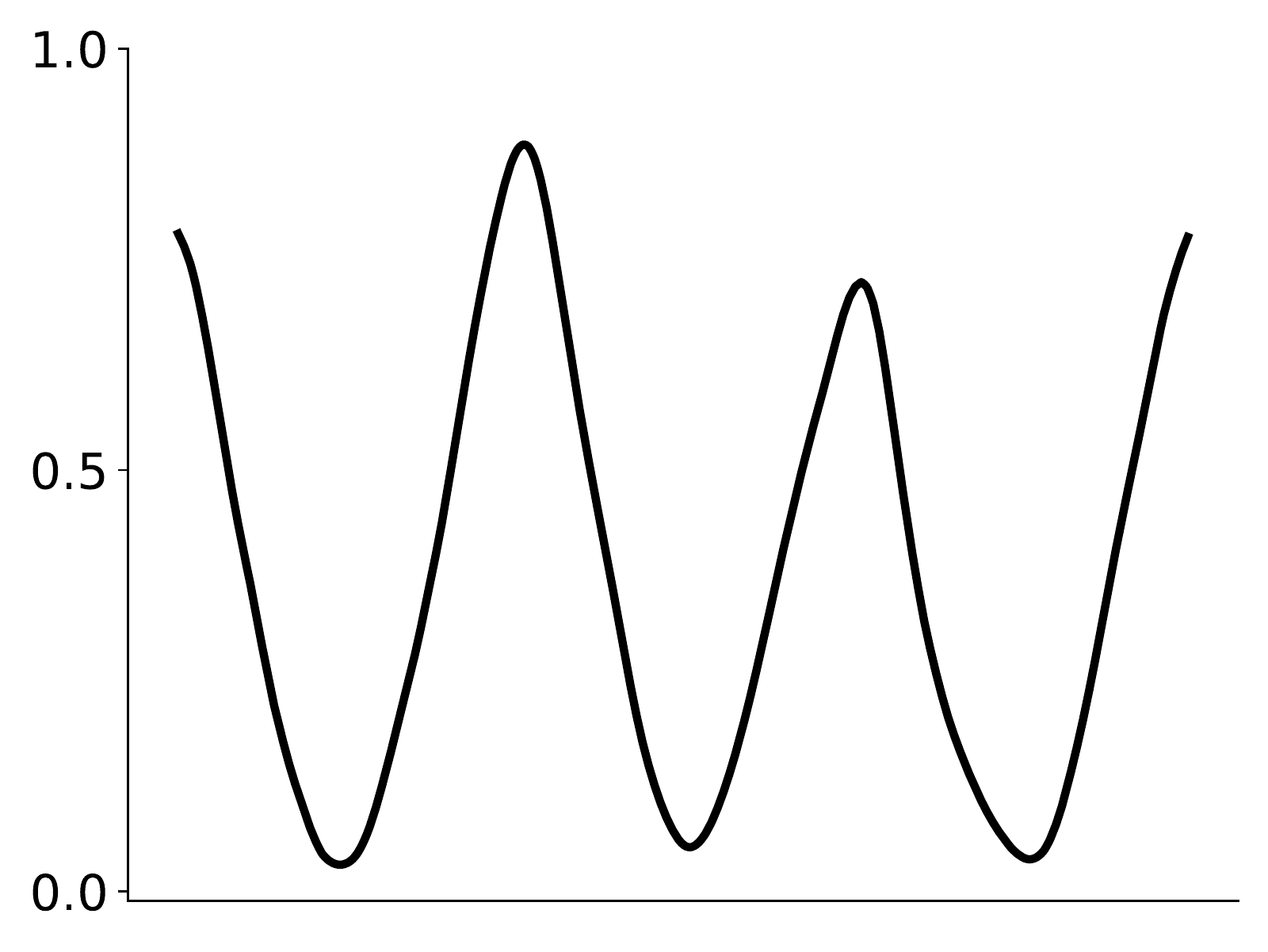}
\end{minipage}
\begin{minipage}{.32\linewidth}
    \centering
    \includegraphics[width=1.0\linewidth]{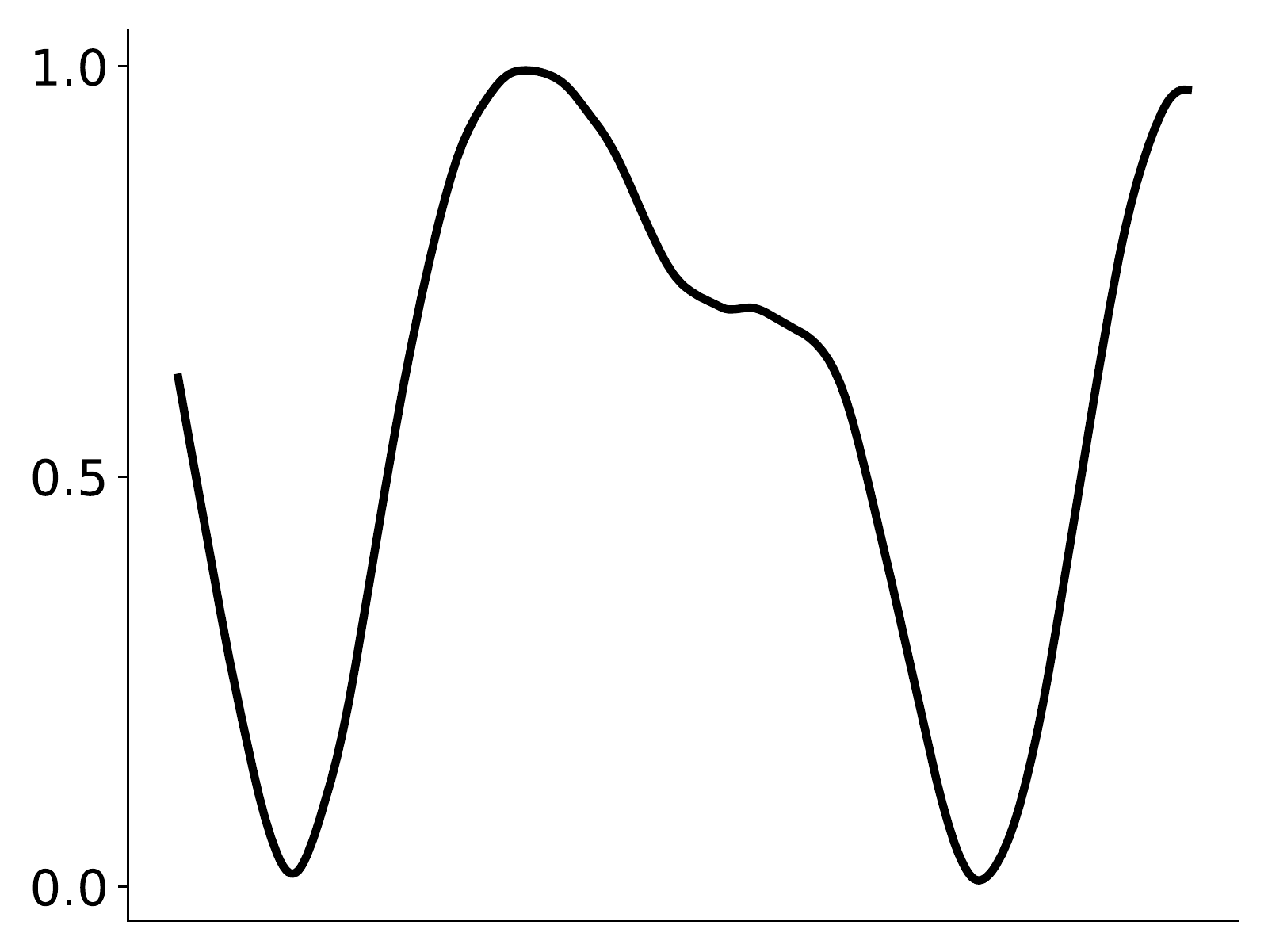}
\end{minipage}
\begin{minipage}{.32\linewidth}
    \centering
    \includegraphics[width=1.0\linewidth]{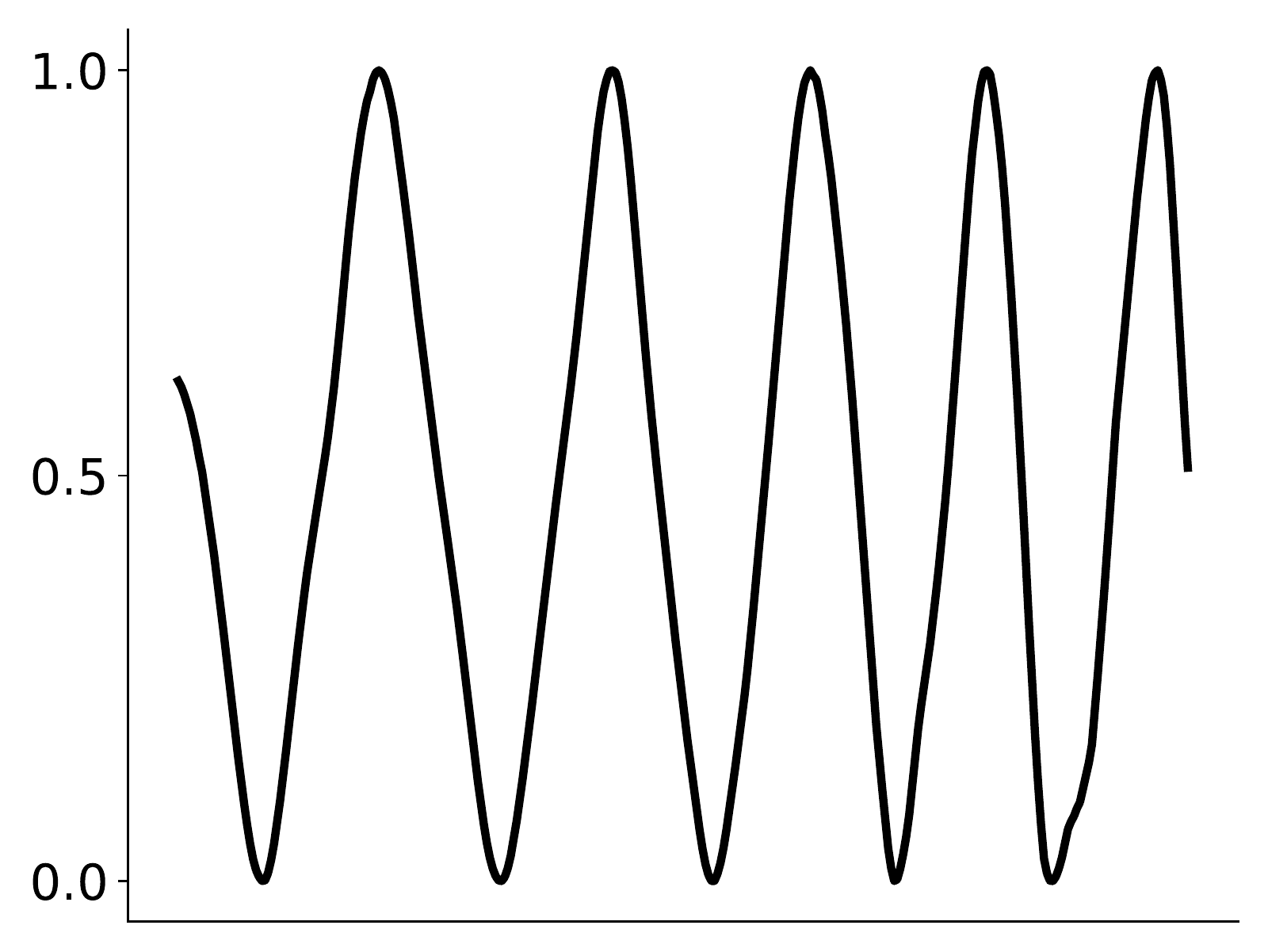}
\end{minipage}
\caption{Extracted LFO patterns from unseen audio effects. \\ Top row: EGFx Phaser, Flanger, Chorus \\ Bottom row: Melda Phaser Irregular, Flanger Irregular, Quasi.}
\label{fig:effect_extractions}
\vspace{-0.4cm}
\end{figure}

We are able to use LFO-net to model unseen analog and digital audio effects using the effect model described in Section \ref{sec:effect_modeling}. Figure \ref{fig:effect_extractions} shows some examples of the extracted LFO signals from the different effects. 
For the EGFx analog effects dataset, we see best results on the phaser effect, followed by the chorus, and then the flanger, which is not able to be modeled effectively. 
We found this dataset to be challenging due to large differences in power supply noise between dry and wet audio pairs, making it difficult to interpret the error metrics and forcing the LSTM to learn to model these differences as well. Despite this, the phaser is able to be modeled and sounds close to the wet audio from informal listening. We provide audio samples in the supplemental material.    

The chorus effect is not modeled very well, but in our initial experiments we found that the LSTM effect model is unable to learn chorus effects, even when presented with the ground truth LFO signal, due to the long delays they make use of. As a result, we are surprised to see that the chorus model performs better than the baseline and is sometimes able to match the volume envelope of the wet audio.  
We also notice that the flanger appears to have two modulations occurring in its spectrogram. LFO-net is able to reliably extract one of them, but this is insufficient for modeling the effect. We believe extracting multiple modulations from audio is a natural future research direction to continue this work on. 

For the Melda digital effects we see that both the irregular and quasiperiodic phaser and flanger effects are able to be captured successfully by the effect model. Our informal listening tests also confirm that they sound close to the target wet audio. The baseline model is able to capture the effects to an extent, but struggles especially with the quasiperiodic and irregular phaser LFO signals. The difference in the final ESR highlights the importance of providing an accurate LFO signal to the effect model.

We plot extracted LFO signals from unseen audio effects in Figure \ref{fig:effect_extractions}. A similar LFO shape to the one shown in Figure \ref{fig:melda_lfo} is extracted for the flanger, but for the phaser it is extracted as two individual rounded peaks, one taller than the other. Since the irregular phaser is able to be modeled with a lower ESR than the irregular flanger, this indicates that this may be an artifact of the internal implementation of the Melda phaser, or that the exact LFO shape may not be required to successfully model an LFO-driven effect. We consider this another interesting future research direction.


\vspace{-4pt}
\section{Conclusions}
\vspace{-4pt}

In this work, we propose a system that extracts arbitrary LFO signals from processed audio for multiple LFO-driven audio effects (phaser, flanger, and chorus), parameter settings, and instrument configurations. 
Our approach does not impose any restrictions on the LFO shape, which allows our neural network architecture to generalize to quasiperiodic, combined, and distorted modulation signals.
We test our pretrained network on LFO extraction from a multitude of unseen audio sources, including guitar, bass, keyboards, drums, and singing voice.
We show through a visualization of the latent space that the network learns meaningful representations of the different modulation shapes, rates, and effects.
Finally, we demonstrate that our pretrained extraction network enables end-to-end modeling of unseen analog and digital LFO-driven audio effects when coupled with a simple processing network, overcoming the need for cumbersome and hand-engineered LFO measurement methods.
We find that asymmetrical and discontinuous LFO shapes, such as saw waveforms, are the most difficult to extract and that the effect model cannot learn LFO-driven effects that make use of larger delays or contain multiple modulations.
We make our code available and provide the trained audio effect models in a real-time VST plugin.

\vspace{-4pt}
\section{Acknowledgments}\label{sec:acknowledgement}
\vspace{-4pt}
Funded by UKRI and EPSRC as part of the ``UKRI CDT in Artificial Intelligence and Music'', under grant EP/S022694/1.

\bibliographystyle{IEEEbib}
\bibliography{DAFx23_tmpl} 

\end{document}